\documentclass{aa}

\usepackage{graphicx}

\usepackage{txfonts}
\usepackage{multicol}
\usepackage{multirow}
\usepackage{hyperref}
\usepackage{ulem}
\usepackage[utf8]{inputenc}
\usepackage{xeCJK}

\begin{document}

   \title{The Redshifts from 122 Bands: Comparative Redshift Forecast for Low-Resolution Spectra from SPHEREx and 7-Dimensional Sky Survey (7DS)}

   \author{Jangho Bae
        \inst{1}
        \and
        Bomee Lee \inst{2}\thanks{Corresponding authors: bomee@kasi.re.kr; myungshin.im@gmail.com}
        \and
        Myungshin Im \inst{1, 3\star}
        \and
        Hyeonguk Bahk \inst{1}
        \and
        Kim Dachan \inst{4}
        \and
        Ho Seong Hwang \inst{1, 3, 4}
        \and
        Sungryong Hong \inst{2}
        \and
        Suk Kim \inst{2}
        \and
        Minjin Kim \inst{6, 7}
        \and
        Taewan Kim \inst{1}
        \and
        Jeyeon Lee \inst{4}
        \and
        Jubee Sohn \inst{1, 3}
        \and
        Hyunmi Song \inst{4} 
        \and
        Seo-Won Chang \inst{1, 3}
        \and
        Yun-Ting Cheng \inst{8}
        \and
        Andreas L. Faisst \inst{9}
        \and
        Zhaoyu Huai \inst{8}
        \and
        Woong-Seob Jeong \inst{2}
        \and
        Ji Hoon Kim \inst{1, 3}
        \and
        Dohyeong Kim \inst{10}
        \and
        Yongjung Kim \inst{11, 12, 2}
        \and
        Seong-Kook Lee \inst{1, 3}
        \and
        Daniel C. Masters \inst{9}
        \and 
        Eunhee Ko \inst{13}}

    \institute{Astronomy Program, Department of Physics and Astronomy, Seoul National University 
    1, Gwanak-ro, Gwanak-gu, Seoul 08826, Republic of Korea \\
    \email{janghobae0604@gmail.com, myungshin.im@gmail.com}
    \and 
        Korea Astronomy and Space Science Institute (KASI) 776, Daedeok-daero, Yuseong-gu, Daejeon, Republic of Korea \\
    \email{bomee@kasi.re.kr}
    \and
        SNU Astronomy Research Center, Seoul National University, Seoul 08826, Republic of Korea
    \and
        Department of Earth, Environmental \& Space Sciences, Chungnam National University 99, Daehak-ro, Yuseong-gu, Daejeon, Republic of Korea
    \and
        Australian Astronomical Optics - Macquarie University, 105 Delhi Road, North Ryde, NSW 2113, Australia
    \and 
        Department of Astronomy and Atmospheric Sciences, Kyungpook National University 80, Daehak-ro, Buk-gu, Daegu, Republic of Korea
    \and
        Department of Astronomy, Yonsei University, 50 Yonsei-ro, Seodaemun-gu, Seoul 03722, Republic of Korea
    \and
        California Institute of Technology, 1200 E California Blvd, Pasadena, CA 91125, USA
    \and
        IPAC, California Institute of Technology, 1200 E California Blvd, 91125, USA, California Institute of Technology, 1200 E California Blvd, Pasadena, CA 91125, USA
    \and
        Department of Earth Sciences, Pusan National University, Busan 46241, Republic of Korea
    \and
        School of Liberal Studies, Sejong University, 209 Neungdong-ro, Gwangjin-Gu, Seoul 05006, Republic of Korea
    \and
        Department of Physics and Astronomy, Sejong University, 209 Neungdong-ro, Gwangjin-Gu, Seoul 05006, Republic of Korea
    \and
        Institut d’Astrophysique de Paris, UMR 7095, CNRS, Sorbonne Université, 98 bis boulevard Arago, F-75014 Paris, France
    }
   \date{Received September XX, XXXX; accepted XX XX, XXXX}

    \abstract
   {The recently initiated SPHEREx and 7DS surveys will deliver low-resolution spectra ($R\approx 30-130$) for hundreds of millions of galaxies over the optical to near-infrared range ($0.4-5.0\,\mu m$), covering a wide sky area without sample selection.
   These unique datasets will improve redshift estimation and provide a rich redshift catalog for the community.
   In this study, we forecast the performance of widely-used photometric redshift estimation methods using simulated SPHEREx and 7DS data. Four template-fitting approaches and two machine-learning (ML) methods are used to derive photometric redshifts from low-resolution spectrophotometric data. We measure redshifts using mock catalogs based on the GAMA and COSMOS galaxy samples and achieve high precision for bright (13 < i < 18) galaxies, with $\sigma_{\textrm{NMAD}}\lesssim 0.005$, bias $\lesssim 0.005$, and a catastrophic failure rate $\lesssim 0.005$ for all methods employed. We find that the combined SPHEREx + 7DS dataset significantly improves redshift estimation compared to using either the SPHEREx or 7DS datasets alone, highlighting the synergy between the two surveys. Moreover, we compare the redshift estimation performance across magnitude ranges for the different methods and examine the probability distribution functions (PDFs) produced by the template-fitting approaches. As a result, we identify some factors that can affect the redshift measurements, like treatments on dust extinction or inclusion of flux uncertainty in the ML model. We also show that the PDFs are relatively well calibrated, although the confidence intervals are generally underestimated, particularly for bright galaxies in the template-fitting methods. This study demonstrates the strong potential of SPHEREx and 7DS to deliver improved redshift measurements from low-resolution spectrophotometric data, underscoring the scientific value of jointly utilizing both datasets.
  }


\titlerunning{Redshift performance forecast of SPHEREx and 7DS}
\maketitle

\section{Introduction}
The redshift of a galaxy is one of the fundamental quantities we can measure through observations.
It is the central quantity for measuring the distance to a galaxy and understanding the 3D structure of the Universe (e.g., \citealp{dawson16_eboss, desi16, reid16, ross20}).
Moreover, the distances obtained from the redshifts scale bulk properties like luminosity and mass, and even correlate with the cosmic epochs of the galaxies.

Spectroscopic observations can measure redshifts robustly, as they can resolve the spectral features such as stellar continua or emission lines of the galaxies \citep{kurtz98, cappellari17, kim25b_rvsnupy}.
However, spectroscopic observations are biased toward the high signal-to-noise targets, like bright galaxies with strong emission lines.
Furthermore, many spectrographs employ slits (e.g. \citealp{hook04}) or fibers (e.g., \citealp{uomoto99, gunn06}) that limit the observable number of targets and necessitate pre-selection of target galaxies (e.g., \citealp{prakash15, comparat16, hahn23, sohn23}). 
These observations are generally more costly and limited to a small number of targets compared to imaging observations.
There are slitless spectrographs \citep{euclid_col22, wang22}, but they can suffer from blending of spectra in a crowded field.

Alternatively, photometric redshift (photo-$z$) estimations from multi-band photometry can provide redshifts of galaxies without spectroscopic data \citep{baum62, koo85, arnouts99, weaver22, feder24, shuntov25}.
Photo-$z$ values have played an important role in the research of galaxy evolution and cosmology (see \citealp{zheng12, salvato19, newman22} for reviews).
Photo-$z$ values are advantageous for studying galaxy populations as they can offer redshifts for all detected galaxies without pre-selection, although they are generally less precise than spectroscopic redshifts due to low spectral resolution ($R = \lambda/\Delta \lambda$) of the photometric data. 

This aspect is particularly beneficial for statistical studies of large galaxy samples. 
For example, when investigating the redshift dependence of galaxy properties with the redshift bins sufficiently larger than the typical error of photo-$z$ estimations, the error of the photo-$z$ becomes insignificant compared to other sources of uncertainty \citep{newman22}.
Furthermore, using statistical measures such as the mean of photo-$z$ values reduces the importance of individual uncertainties, thereby improving the precision in science cases that rely on the statistical properties of large samples.
Therefore, many studies for probing large-scale structures (e.g., \citealp{boris07, sanchez11, scoville13, ko24, euclidcol25}), galaxy clustering (e.g., \citealp{durret11, carnero12, castignani16, crocce16, zhou21}), weak and strong lensing (e.g., \citealp{treu10, mandelbaum18, cha25}), and cosmology (e.g., \citealp{blake05, seo12, masters15, hildebrandt17}) utilized photo-$z$ values.

We can estimate the photo-$z$ values of the galaxies with two primary approaches. 
The first one is a template-fitting method that exploits the prior knowledge of the spectral energy distributions (SEDs) of galaxies \citep{arnouts99, brammer08, stickley16, lee20, brammer21, laur22_topz}.
This method compares the observed SED from multi-band photometry with the template SEDs and finds the best-fit template and redshift.

The second one is inferring the relation between the photometric data and the true redshifts using machine learning (ML) techniques \citep{collister04_annz, beck16, laur22_topz, kim25a, pathi25_annzp}.
These ML-based methods try to find the relation between the multi-band data and the true redshifts. 
They generally train the relation using a training sample with known redshifts. 
In the literature, many ML techniques like neural network \citep{collister04_annz, pathi25_annzp}, random forest (RF, e.g., \citealp{Carrasco_kind13, kim25a}), convolutional neural network (e.g., \citealp{henghes22}), and Gaussian process (e.g., \citealp{gomes16, almosallam16}) are utilized to measure photo-$z$ values. 

Although these efforts enhance the photo-z precision and accuracy above $1/R$, the intrinsically low R of broadband data limits photo-z precision and accuracy for the given depth.
This limits the applicability of photo-$z$ values in cosmological and galactic studies.
For example, scatters in the photo-$z$ values can obscure the large-scale structure of galaxies, while biases and catastrophic failures can introduce biases in clustering analyses \citep{newman22}.

The low-resolution spectral surveys utilizing medium- to narrow-band filters can yield improved photo-$z$ values through denser spectral sampling compared to broadband surveys.
\cite{wolf03} pioneered medium-band surveys with the Classifying Objects by Medium-Band Observations in 17 Filters (COMBO-17) survey.
They showed redshift accuracy of $\sigma \approx 0.03$ for galaxies with $R$ magnitude $\lesssim 24$.
MUSYC (Multi-wavelength Survey by Yale-Chile, \citealp{taylor09, cardamone10}) used 18 optical medium bands and other ancillary data and yielded $\sigma\sim0.008$ for galaxies with $R$ magnitude $< 25.3$.
The Physics of the Accelerating Universe Survey (PAUS, \citealp{eriksen19}) utilized 40 narrow-band filters covering $4500-8500$\r{A} wavelength range.
They achieved $\sigma /(1+z) = 00037$ for the galaxies with magnitude down to $i\sim22.5$ and with upper 50\,\% photo-$z$ qualities.
More recently, the Javalambre-Physics of the Accelerating Universe Astrophysical Survey (J-PAS, \citealp{benitez14}) plans to survey $8500\,\deg^2$ of Northern sky with 54 narrow-band filters by utilizing a wide-field telescope with multiple detectors.

The recently started SPHEREx (Spectro-Photometer for the History of the Universe, Epoch of Reionization, and Ices Explorer, \citealp{dore14, dore16, dore18, crill20, spherex25}) and 7DS (7-Dimensional Sky Survey, \citealp{kim24_7dt}) projects are planned to overcome the limitation from coarse spectral sampling of broadband surveys across a wide sky area.
SPHEREx enhances sky coverage by combining wide-field optics and linear variable filters (LVFs).
7DS improves survey efficiency through simultaneous observations using multiple wide-field telescopes equipped with different medium-band filters.
These surveys will offer low-resolution ($R\sim 20 - 130$) spectra for all the objects in the field of view from optical (7DS) to NIR (SPHEREx) wavelength range ($0.4 - 5\mu$m) in a wide sky area.

The large volume of low-resolution spectral data from SPHEREx and 7DS will enable more precise and accurate redshift measurements.
The SPHEREx and 7DS data lie between spectroscopic and photometric surveys, making them valuable for both spectroscopic and photometric redshift measurements.
In this study, as we use the methods that have been utilized to measure photo-$z$s mainly exploiting the continuum features, we will refer to the redshifts we obtain as photo-$z$.

Some previous studies assessed the performance of photo-$z$ estimation with the simulated galaxy catalog of SPHEREx and 7DS.
\cite{stickley16} and \cite{feder24} tested the photo-$z$ estimation performance from mock SPHEREx data using the SPHEREx in-house code for photo-$z$ measurements.
The photo-$z$ values from \cite{feder24} showed $\sigma \sim 0.0025$, $b$ (bias) $\sim -0.0001$, and $\eta$ (catastrophic failure rate) $\sim 0.02$ for the brightest galaxies in the sample ($16.5 < W1 < 18.5$).

\cite{ko25} tested the photo-$z$ estimation performance with mock 7DS data using EAZY \citep{brammer08}.
They reported $\sigma \sim 0.003 - 0.007$, and $\eta \sim 0.008 - 0.081$ for galaxies with $19 \leq m(\lambda=625\rm{nm}) < 22$.
They also tested a combination of 7DS data with mock Pan-STARRS1 \citep{chambers16}, VIKING \citep{edge13_viking}, and SPHEREx full-sky survey data.
In these tests, they find significant enhancement in the performance, especially when combined with SPHEREx full-sky survey, achieving $\sigma=0.003-0.006$ and $\eta = 0.000-0.004$ for the same magnitude range.

These studies demonstrated the potential of SPHEREx and 7DS for providing precise photo-$z$ values for a large number of galaxies.
However, \cite{feder24} mainly evaluated photo-$z$ performance using mock SPHEREx and \textit{g}, \textit{r}, \textit{z}, W1, and W2 data with SPHEREx in-house photo-$z$ code.
On the other hand, \cite{ko25} conducted a photo-$z$ performance assessment for combined SPHEREx + 7DS data, but they tested the performance solely with EAZY \citep{brammer08} and employed the survey parameters for both SPHEREx and 7DS before updates.
Since each study utilized one photo-$z$ code, it remains unclear whether the photo-$z$ performance of SPHEREx and 7DS data is consistent across different photo-$z$ estimation codes.
Furthermore, the synergy between these surveys can improve the scientific outputs beyond the intended level of the survey design, but this aspect was not fully addressed with different photo-$z$ measuring strategies before.

In this paper, we evaluate the photo-$z$ performance of SPHEREx and 7DS surveys by applying six different photo-$z$ estimation methods to the simulated survey datasets.
We aim to assess the potential and the synergy of these surveys for photo-$z$ measurements.
We also attempt to investigate the strengths and weaknesses of different methods for this unique data and find more optimized settings for future studies.

In Section \ref{sec:data}, we outline the SPHEREx and 7DS surveys and the procedure for generating SPHEREx and 7DS mock catalogs. 
We describe the methods for estimating photo-$z$ values in Section \ref{sec:method}.
We present the results for SPHEREx + 7DS, SPHEREx, and 7DS datasets in Section \ref{sec:result} with the comparison between the methods as a function of $i$-band magnitude and true redshift.
In Section \ref{sec:discussion}, we discuss the photo-$z$ performance of SPHEREx and 7DS data and the synergy between them. We also analyze the differences in the results from different methods.
Then we summarize our results and conclude in Section \ref{sec:conclusion}.

Throughout this paper, we will use the AB magnitude system \citep{oke83_AB}.

\section{Data} \label{sec:data}
In this section, we outline the survey instruments, survey plans, and their scientific potential.
We also summarize the SED templates and survey data for generating a mock catalog.
We then describe the procedure to construct mock SPHEREx and 7DS galaxy catalogs that we use for estimating photo-$z$ performance of the surveys.

\subsection{Instruments \& Surveys}
\begin{figure}[h!]
\includegraphics[width=0.47\textwidth]{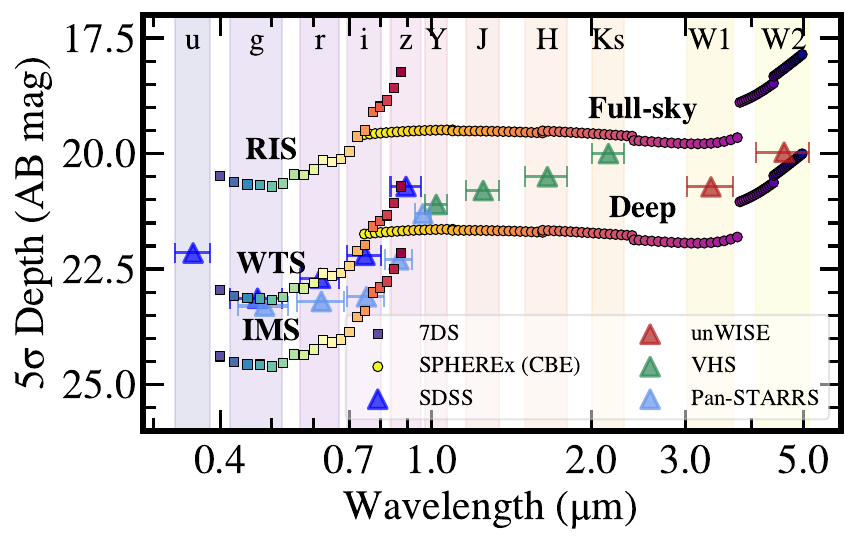} 
\caption{The 5$\sigma$ point source detection limits of SPHEREx surveys and 7DS. The colored squares show the depths of the three layers of 7DS, with the RIS at the top, followed by the WTS and IMS layers. The circular markers show the depths of SPHEREx full-sky and deep surveys from the top. The blue, red, green, and pale blue triangles denote the 5$\sigma$ point source depths of other popular optical and NIR imaging surveys (SDSS \citep{york00_sdss}, unWISE \citep{schlafly19_unwise}, VHS \citep{mcmahon13_vhs}, and Pan-STARRS \citep{chambers16}). 
The shaded regions show the wavelength ranges of the plotted broadband filters (\textit{u, g, r, i, z, Y, J, H, $K_s$}, W1, and W2) with the names indicated at the top.} 
\label{fig:depth}
\end{figure}

\begin{table} 
\caption{Summary of SPHEREx and 7DS surveys. \label{tab:survey_info}}
\centering 
\begin{tabular}{c|c|p{2.5cm}|p{1.7cm}}
\hline
Survey & Layer & Depth ($m_{AB}$) & Area ($\deg^2$) \\
\hline\hline
\multirow{4}{*}{SPHEREx} & \multirow{2}{*}{Full-sky}& $17.5$ ($2\,\mu m$) -- $19.5$ ($5\,\mu m$) & \multirow{2}{*}{Full sky}\\  
{       }                & \multirow{2}{*}{Deep}    & $19.5$ ($2\,\mu m$) -- $21.5$ ($5\,\mu m$) & \multirow{2}{*}{$\sim200$}\\
\hline
\multirow{6}{*}{7DS}     & \multirow{2}{*}{RIS}     & $18.2$ ($0.5\,\mu m$) -- $20.7$ ($0.9\,\mu m$) & \multirow{2}{*}{$\sim24,000$}\\ 
{       }                & \multirow{2}{*}{WTS}     & $20.7$ ($0.5\,\mu m$) -- $23.2$ ($0.9\,\mu m$) & \multirow{2}{*}{$\sim1,600$}\\
{       }                & \multirow{2}{*}{IMS}     & $22.1$ ($0.5\,\mu m$) -- $24.6$ ($0.9\,\mu m$) & \multirow{2}{*}{$\sim10$}\\
\hline
\end{tabular}
\tablefoot{The depths of the survey layers depend on the wavelengths.}
\end{table}

Figure \ref{fig:depth} shows the depths of different survey layers, along with some other established large surveys with similar wavelength ranges.
The figure shows the different depths of the three layers of 7DS and the two layers of SPHEREx surveys with square and circular markers, respectively.
All the depths plotted are 5$\sigma$ point source depths.
The current best estimates (CBE) depths of SPHEREx surveys are based on the SPHEREx public git repository \footnote{github.com/SPHEREx/Public-products}.
We simulate the depths of 7DS assuming the same observational and environmental parameters we use for generating the mock 7DS catalog used in this study, which is detailed in Section \ref{sec:mock_generation} and \cite{ko25}.
Figure \ref{fig:depth} also shows the survey depths of SDSS \citep{york00_sdss}, Pan-STARRS 3$\pi$ survey \citep{chambers16}, VHS\footnote{Depths from www.eso.org/rm/api/v1/public/releaseDescriptions/144} \citep{mcmahon13_vhs}, and unWISE 
\citep{schlafly19_unwise}.
We also summarize the depths and survey coverage of the SPHEREx surveys and 7DS in Table \ref{tab:survey_info}.

\subsubsection{SPHEREx}
SPHEREx is a NASA Medium Explorer (MIDEX) mission that was successfully launched on March 11, 2025. It began its 25-month all-sky survey following completion of in-orbit checkout on May 1, 2025 \citep{dore14, crill20, spherex25}. SPHEREx is conducting the first full-sky spectrophotometric survey in the NIR at wavelengths from 0.75 to 5.0$\,\mu$m. Using a 20\,cm wide-field telescope with a $3.5 \times 11.3\,\deg ^2$ instantaneous field of view, SPHEREx obtains spectra for every $6\farcs15\times6\farcs15$ pixel on the sky with resolving powers ranging from $R=35-130$ \citep{korngut18}. SPHEREx utilizes LVFs, whose central wavelength varies linearly with position across the field of view. This unique technique allows SPHEREx to build a complete spectrum for each object of interest through successive exposures. Over its nominal two-year mission, SPHEREx will perform four all-sky surveys, capturing data across 102 spectral channels per sky position. To achieve Nyquist sampling of the spectral response, observations are offset by half a spectral channel between alternating survey passes. 

SPHEREx’s survey strategy includes both a Full-Sky Survey and a Deep Survey. Each point along the ecliptic will be observed at least four times, with significantly higher redundancy in the deep survey fields near the North and South Ecliptic Poles (NEP and SEP), covering about 200 square degrees. Figure \ref{fig:depth} shows the 5$\sigma$ sensitivity per spectral channel, derived from the CBE. For the full-sky survey, the sensitivity reaches $m_{AB} \sim$ 19.5 from 0.75 to 3.8$\,\mu$m and $m_{AB} \sim$ 19$-$17.5 from 3.8 to 5$\,\mu$m. The deep survey regions achieve sensitivities that are approximately 2 magnitudes deeper than those of the full-sky survey.

Over its two-year primary mission, SPHEREx will provide a rich legacy catalog of over 1 billion galaxy spectra, over 100 million high-quality galaxy redshifts, tens of millions of stellar and ice absorption spectra, and thousands of quasar and asteroid spectra\footnote{The mission's data products will be publicly accessible through NASA's Infrared Science Archive (IRSA).}. Its scientific goals include mapping the 3D large-scale structure of the universe up to redshift $z \lesssim 2$ to constrain inflationary physics, conducting intensity mapping of star formation history, and tracing the cosmic journey of water and biogenic molecules from the interstellar medium to planetary systems.

\subsubsection{7DT \& 7DS}
7DT \citep{kim24_7dt} is an array of wide-field telescopes located in Chile's El Sauce observatory, consisting of 20 unit telescopes.
Each unit telescope has a wide field of view of $1.33\times 0.89 \deg^2$ with a 50\,cm aperture and is equipped with a CMOS camera with a pixel scale of $0\farcs5$. 

7DT can observe the wide field of view using medium-band filters with 25\,nm bandwidths and central wavelengths from 0.4 to 0.9$\,\mu$m ($R\sim 30-60$), with the central wavelengths between the adjacent filters separated by 12.5$\,$nm.
The central wavelength of the bluest medium-band filter is 400$\,$nm, while it is 875$\,$nm for the reddest one.
The filters are named such as $m400$, where $400$ is the three-digit number of the central wavelength in nm, while $m$ stands for medium-band.

When each of 20 7DT telescopes observes the same field using two medium-band filters with different central wavelengths, we can obtain a 40-wavelength spectral image over the entire field of view.
This unique instrumental design will address various science cases, such as galaxy evolution \citep{kim24_7dt, lim25}, transient search (\citealp{paek24, khalouei25}), exoplanet atmospheres (Bae et al., in preparation), and cosmology. 

Currently, 7DT uses 20 medium-band filters with each filter separated by 25nm in central wavelengths ($m400, m425, ..., m850$, and $m875$).
Throughout this paper, we assume 7DT observations with these currently installed 20 filters.
We expect that the future installation of an additional 20 filters would improve photo-$z$ estimation accuracy (c.f. \citealp{ko25}).

7DS is an optical spectro-photometric survey conducted with 7DT \citep{kim24_7dt}, utilizing medium-band filters.
7DS consists of three layers with different survey depths and sky coverages to address a wide range of sciences: Reference Image Survey (RIS), Wide-field Time-domain Survey (WTS), and Intensive Monitoring Survey (IMS).

RIS will survey nearly the entire sky at Dec $< 20\,\deg$ observable from the 7DT at the El Sauce observatory, Chile (30.4725$^{\circ}$ S, 70.7631$^{\circ}$ W).
The survey will capture the images with 300 seconds of on-source integration time per filter and tile.
This survey will provide the widest view of the sky among the three survey layers and will be used as reference images for transient search with target-of-opportunity observations.
This survey is currently running, and we expect to complete the survey by next year.

WTS will visit each survey tile every $\sim20$days with the same exposure time as the RIS per visit.
This survey will cover $\sim1,600\,\deg^2$ of the Southern sky. 
The survey field has not been decided, but WTS will explore long-term spectral variation of various objects such as active galactic nuclei (AGN).
With 5 years of observations, this survey will have comparable depths to the SDSS survey ($m_{AB}\lesssim 23$, see Figure \ref{fig:depth}) with medium-band filters. 

IMS will provide the deepest data among the 7DS layers. 
7DT will visit the IMS survey tiles every observable night, providing the deepest data in the 7DS.
The survey area will overlap the SEP field of the SPHEREx deep survey to maximize the synergy between the two deep surveys.

\subsection{Mock Catalog}
The mock catalog of SPHEREx and 7DS data is from the simulated SEDs from \cite{feder24}, who constructed mock spectra of the galaxies based on the COSMOS2020 \citep{weaver22} and GAMA \citep{driver22} multiband photometry data. 
Here we outline the galaxy SED templates and introduce the COSMOS and GAMA data used in this study.
After that, we summarize the procedure to make mock catalogs used to test the performance of photo-$z$ estimation.

Although we try to faithfully simulate the real observations by considering several observational factors, the mock catalog of SPHEREx and 7DS will have some differences compared to real data.
The environmental and instrumental effects may be more complicated than our assumptions.
In spite of that, our results can serve as a benchmark of the photo-$z$ measurement performance of SPHEREx and 7DS data using different methods.
Furthermore, the homogeneously processed mock data present an ideal sample for comparing the photo-$z$ performance of different combinations of survey layers and different photo-$z$ methods.

\subsubsection{COSMOS and GAMA Data}
A large sample of galaxies with ground-truth redshifts is a key to accomplishing a robust assessment of photo-$z$ measuring performance. 
COSMOS data is ideal for this because it provides a deep and complete catalog of galaxies, representing the realistic distribution of magnitudes and redshifts of galaxies in the COSMOS field.
It is based on the multiband photometry from COSMOS2020 \citep{weaver22}.
The COSMOS2020 data consists of very deep ($i \lesssim 27$) photometric data from 0.15$\,\mu$m to 8$\,\mu$m, covering $\sim 2\,\deg^2$ of the COSMOS field \citep{scoville07, weaver22}.

Among the COSMOS2020 galaxies, \cite{feder24} selected galaxies with $i < 25$ and having at least one NIR band (i.e., UltraVISTA \textit{J} or \textit{H} band). 
They further selected galaxies with robust photo-$z$ estimates that have consistent photo-$z$ values from the two different photometry sets in \cite{weaver22}.
These selection criteria leave 166,014 galaxies in the COSMOS field within the magnitude range of $18<i<25$.

The COSMOS dataset provides a deep sample of galaxies well beyond the limiting magnitude of SPHEREx and 7DS surveys (see Figure \ref{fig:depth}).
However, it has a relatively low number of bright galaxies, which poses difficulty in assessing the photo-$z$ performance for bright and low-redshift galaxies with $i \lesssim 20$.
As the SPHEREx full-sky, WTS, and RIS surveys cover a wide area with relatively shallow depths, measurements of photo-$z$ performance for bright galaxies are important to assess their scientific potential.

In this regard, we also use the mock SEDs of GAMA galaxies \citep{driver22}, which were also constructed by \cite{feder24} from the multiband photometry included in the GAMA dataset.
The GAMA sample contains 44,135 galaxies with $i\lesssim 18.5$ across the four GAMA fields spanning $\sim 200\deg^2$ in total \citep{feder24}.

Figure \ref{fig:iz_dist} shows the distribution of GAMA and COSMOS samples in $i-z_{\rm{true}}$ parameter space. 
The GAMA and COSMOS samples show distinctive distributions in magnitudes and true redshifts.

\subsubsection{Brown+COSMOS Templates}
We use Brown+COSMOS templates for template fitting methods, which are compiled by \cite{feder24} and also used for constructing mock SEDs by fitting the observed multi-band photometry data with these templates (See \ref{sec:data_mockseds} and \citealp{feder24} for more details).
They combined template spectra from \cite{brown14} (Brown template) and \cite{ilbert09} (COSMOS template).
The number of the compiled template set used in \cite{feder24} is 160, which consists of 129 Brown templates and 31 COSMOS templates.

Brown templates are a collection of local galaxy SEDs compiled and extended in \cite{brown14}.
This atlas of galaxies spans a broad range of galaxy types and environments. 
They combined optical and IR spectra and filled the wavelength range without observed spectra based on the SED model fitted to the photometric data using \texttt{MAGPHYS} model \citep{de_cunha08}.
The resulting templates have wavelength coverage from 100\r{A} to 350,000\r{A}.
They represent the broad population of local galaxies.

The COSMOS templates are a collection of templates from \cite{polletta07} and 12 additional model templates generated using the BC03 \citep{bruzual03} stellar library for both starburst and passive elliptical galaxies.
This template set has been widely used in COSMOS surveys \citep{ilbert09, ilbert13, laigle16, weaver22}.
COSMOS templates complement Brown templates by including model-based templates and templates based on higher-redshift galaxies compared to \cite{brown14}.

\subsubsection{Mock SEDs}\label{sec:data_mockseds}
We adopt the mock SEDs constructed by \cite{feder24}.
We will briefly describe the procedure for making mock SEDs used in this study, which is explained in \cite{feder24} with more detail.
They generated mock SEDs using SED fitting to the observed COSMOS and GAMA multiband photometric data with \texttt{Fitcat} as in \cite{stickley16}.
Based on the photo-$z$ value and the multiband photometric data, they modeled the SEDs with the Brown + COSMOS galaxy templates by finding the best-fit template, $E(B-V)$, dust law, and stellar mass.
\cite{feder24} report a median $\chi^2_r$ value of 2.2, with fewer than 1\,\% of galaxies showing $\chi^2_r > 10$. 
The relatively high $\chi_r^2$ values are partly driven by the limited template set, which especially may not fully capture complex PAH features, and by potentially underestimated photometric uncertainties in the COSMOS and GAMA photometric data.
Despite these limitations, the errors in the mock SEDs are generally no larger than the expected photometric uncertainties of the SPHEREx and 7DS surveys, given the depths of the COSMOS and GAMA datasets.
Although the IMS data reach similar depths, the mock SEDs from \cite{feder24} remain useful for providing realistic SEDs and enabling direct comparisons to shallower datasets.
However, some template-driven biases or residuals may become detectable in the deepest IMS data.

Additionally, \cite{feder24} augmented emission lines to the fitted spectra. 
The strengths of H$\alpha$ and [O II] were determined from the best-fit templates, and the observed scaling relations were then used to estimate the [N II] and H$\beta$ line strengths, which were incorporated into the mock spectra. 
They also tested the injected emission-line strengths by comparing the line equivalent widths and luminosity functions with external measurements of COSMOS galaxies, and reported overall consistency.

\subsubsection{Mock SPHEREx and 7DS Data} \label{sec:mock_generation}
We use the mock SEDs from \cite{feder24} to generate mock catalogs of SPHEREx and 7DS by simulating the observations. The SPHEREx catalog is generated using the SPHEREx Sky Simulator \citep{crill25}, a software tool designed to simulate the infrared observations of the SPHEREx mission. The simulator produces realistic infrared sky scenes by astrophysical models (such as zodiacal light, diffuse galactic light, etc), the instrument’s unique spectrometer design and custom onboard detector readout, as well as the SPHEREx all-sky survey strategy. It incorporates realistic noise and systematic effects derived from up-to-date astrophysical measurements and pre-launch instrument characterization campaigns. The simulator can generate full spectral images per detector, matching the layout of the six detectors in one exposure, and produce photometric catalogs by performing forced photometry on the spectral image at pre-selected sky positions. 

The SPHEREx Sky Simulator includes a QuickCatalog mode that bypasses full spectral image generation and directly simulates photometric measurements for given sources over the SPHEREx mission. We use this mode to generate simulated SPHEREx photometry for sources from the COSMOS2020 and GAMA catalogs. The inputs to the simulator are the mock SEDs and sky positions of the sources. Using the SPHEREx survey plan, the simulator identifies the pointings that observe each source, identifies detector positions, and computes the associated bandpasses. 
The transmission curves of the 102 bandpasses were generated from the wavelength response of the SPHEREx arrays by finding all pixels with the pre-defined central wavelengths.
Simulated observations include realistic background emission, simulated point spread functions, and lab-measured noise realizations to enable accurate modeling of photometric uncertainties.

The generation of 7DS mock data generally follows the procedure detailed in \cite{ko25}.
We first convolve the mock SEDs with the 7DS filter transmission curves.
The 7DS filter transmission curves contain the information of the filter transmittance, quantum efficiency of the detector, telescope throughput, and sky transmission.
For the detector and telescope information, we use instrumental information of C3-61000 PRO camera\footnote{https://www.gxccd.com/art?id=647\&lang=409} and PlaneWave DeltaRho 500 (DR500) telescope\footnote{https://planewave.com/products/deltarho-500-ota/} \citep{kim24_7dt}.
For sky transmission, we use the atmospheric model of Cerro Paranal site \citep{noll12}, with an airmass of 1.3 and a precipitable water vapor (PWV) of 2.5mm, following \cite{ko25}.

We calculate the error of the convolved fluxes by considering the Poisson noise, readout noise, and calibration error. 
We assumed g2750 gain mode with a typical gain of 0.26$e^-$/ADU. 
This gain mode is selected for the 7DS observations due to its lower readout noise (3.51$e^-$ for g0 v.s. 1.46$e^-$ for g2750) and thus enables the detection of faint sources.
We assumed 2" seeing, 2"-radius aperture for photometry, an observable night fraction of 70\,\%, and a calibration error of 1\,\% of the flux values.
Using the simulated flux and the errors, we resampled the flux values with Gaussian distributions to mimic observed fluxes.

Although we follow the method for mock data generation, the process has some differences compared to \cite{ko25}.
We use the mock SEDs from \cite{feder24}, but \cite{ko25} used EL-COSMOS mock SEDs \citep{saito20}. 
Also, we apply updated survey strategies with different exposure times, filters, and detector gain settings.
In particular, we assume the current 20 medium-band filters for the 7DT filter set-up, while \cite{ko25} focused more on the 7DT performance using 40 medium-filters.

We define the deep and wide datasets considering the survey area and potential synergy between the survey layers of SPHEREx and 7DS. 
We will focus on these two datasets in the following sections. 
We also present a 7DT RIS simulated sample as a reference for photo-z performance in comparison to other datasets. 

\begin{figure}[h!]
\includegraphics[width=0.47\textwidth]{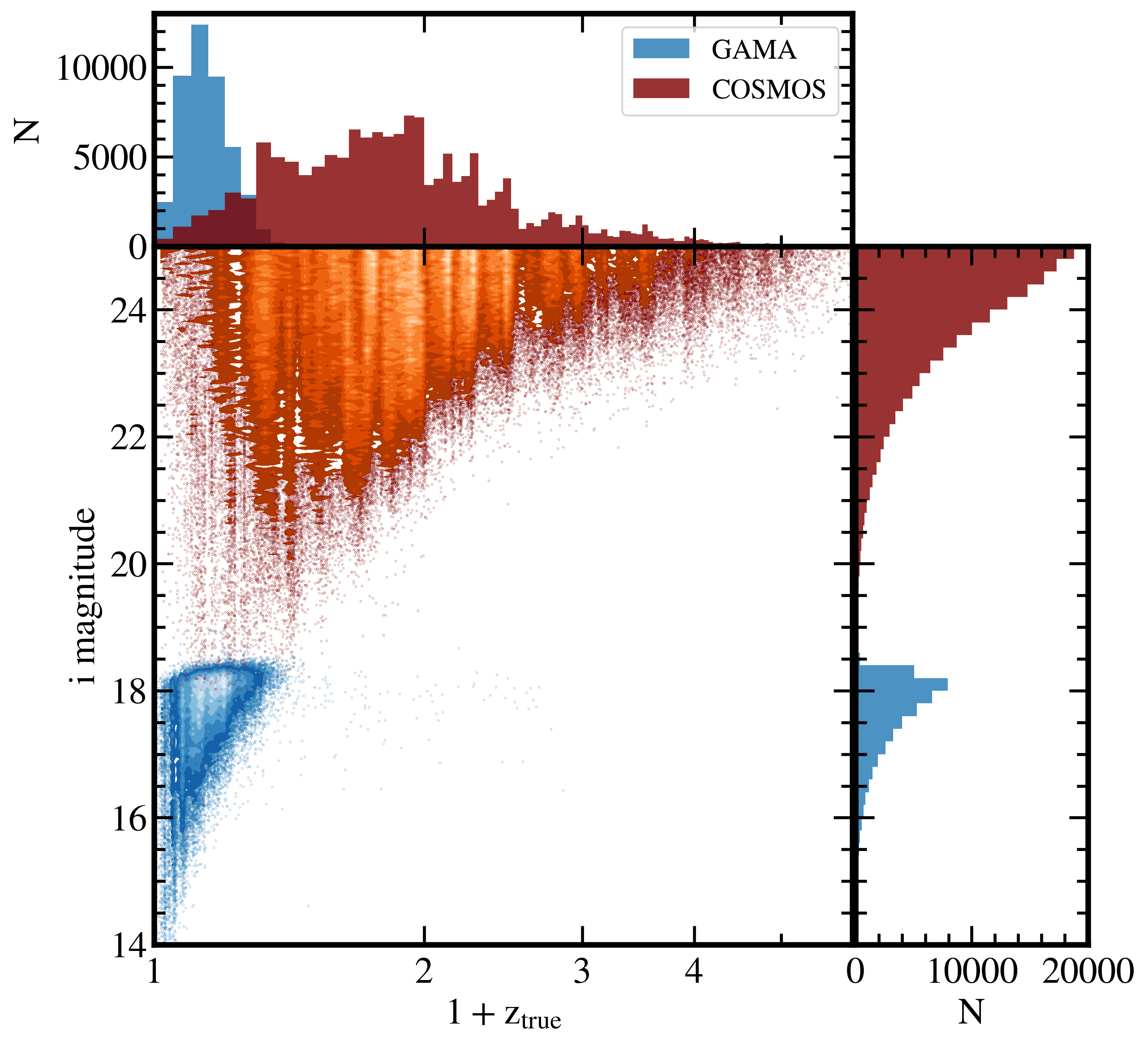}
\caption{The redshift and $i$ magnitude distribution of GAMA and COSMOS galaxies used in this work. The GAMA data (bright blue) consists of brighter galaxies ($i \lesssim 18.5$) with low redshifts ($z \lesssim 0.4$) compared to COSMOS (maroon) galaxies with $18 \lesssim i \lesssim 25$ and $z \lesssim 6$.
\label{fig:iz_dist}}
\end{figure}

The wide dataset contains WTS from 7DS and all-sky survey from the SPHEREx survey.
This dataset includes all the COSMOS and GAMA galaxies, totaling $\sim 210,000$ galaxies with $13\lesssim i\lesssim 25$ (c.f. Figure \ref{fig:iz_dist}).
This sample will be used to test the surveys with wide sky coverages and relatively shallower depths.

The deep dataset consists of IMS from 7DS and the deep survey from SPHEREx. 
As these two datasets cover relatively smaller sky areas intensively with more visits, this deep dataset is suitable for testing photo-$z$ performance for fainter galaxies compared to the wide dataset.
We randomly select 1,000 galaxies within the COSMOS dataset to make a representative sample of faint galaxies and define them as the deep sample.

\begin{figure}[h!]
\includegraphics[width=0.47\textwidth]{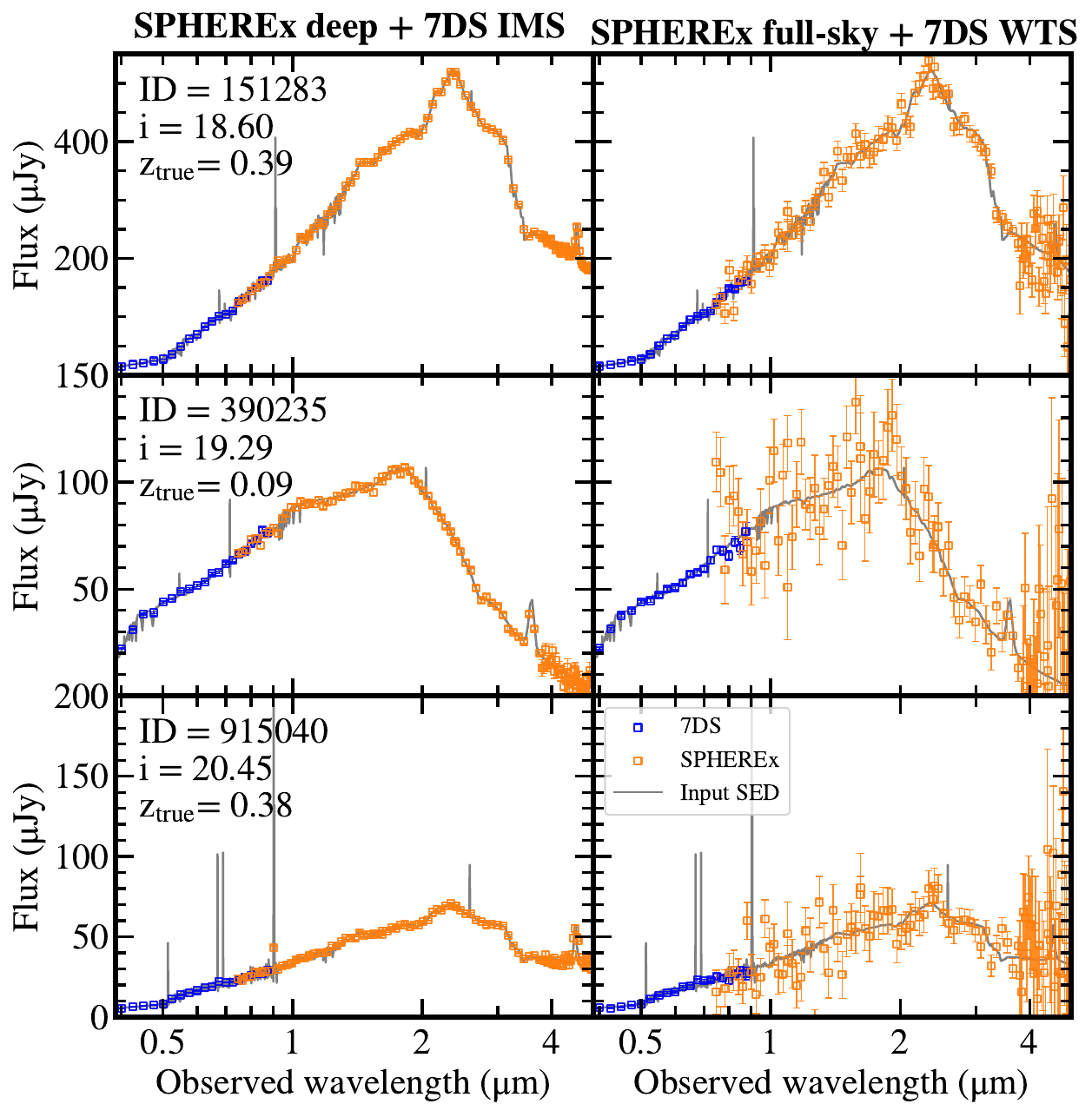}
\caption{Simulated SPHEREx and 7DS data used in this study. The grey plot shows the mock SED from \cite{feder24}. 
The left column of the figure shows the simulated photometry of the deep dataset (SPHEREx deep + 7DS IMS) and the right column shows that of the wide dataset (SPHEREx full-sky + 7DS WTS), with the same example galaxies sorted in ascending $i$ magnitudes. 
The blue markers show the mock datapoints from the 7DS, and the orange markers show the simulated SPHEREx datapoints.
\label{fig:sed}}
\end{figure}

Figure \ref{fig:sed} shows the simulated mock SEDs from the SPHEREx and 7DS overplotted on the original SEDs from \cite{feder24}.
From the top to the bottom rows, the magnitudes of the sample galaxies become fainter.
The simulated photometry reveals the differences in signal-to-noise ratio (S/N) between the wide and deep datasets. 
Specifically, continuum features such as the 4000\,\r{A}-break and the 1.6$\,\mu$m bump are well captured in our mock data.
The broad emission feature from PAH at 3.3$\mu$m is also captured for the galaxies at low redshifts.
Also, some strong line features like H$\alpha$ are captured in the data, as noted in previous research utilizing SPHEREx and 7DS mock data \citep{feder24, ko25}.

The mock catalogs contain some negative flux values resulting from random sampling of fluxes with their corresponding uncertainties.
We treat these negative fluxes as non-detections and either exclude them or substitute them with dummy values during photo-$z$ measurements (see Section \ref{sec:method}).
To remove the dependence of the training and test sets on the results of different ML-based methods, we define the training and test sets in advance with \texttt{FLAG\_ML}. 
The 20\,\% of the galaxies in our sample are randomly selected as the test sample, and we assign them $\texttt{FLAG\_ML} = 1$.

The COSMOS galaxies with $i \sim 25$ are much fainter than the survey limit of the wide dataset.
Therefore, we used the COSMOS galaxies with $18<i<21$ ($N=5,065$) for presenting the results from the wide dataset, as this magnitude range covers up to the median depth of the wide dataset (see Figure \ref{fig:depth}).
We use fainter samples up to $i=23$ when we present and compare the results from the deep dataset.
Though we do not use the whole sample of COSMOS galaxies for presenting results, we use it for training the models of ML-based methods.
Note that when we show the results from ML-based methods, we present only the results from the test sample ($\texttt{FLAG\_ML}=1$).

\section{Photo-$z$ Methods} \label{sec:method}
We independently test and optimize 6 methods to evaluate the performance for photo-$z$ estimation of SPHEREx and 7DS mock data.
Using these independent methods and configurations, we aim to demonstrate that SPHEREx and 7DS data consistently yield accurate photo-$z$ estimates.
We also investigate some factors that influence the performance of photo-$z$ estimation with these approaches.
In this section, we outline the methods and settings for measuring photo-$z$ values.
We test four template-fitting methods and two ML-based methods.

\subsection{SPHEREx in-house pipeline} \label{sec:method_inhouse}
SPHEREx in-house is a photometric redshift estimation code implemented in \citet{stickley16} and \citet{feder24}. It is fundamentally similar to the widely used template-fitting code LePhare \citep{arnouts99, ilbert06}, but has been re-implemented in C++ and adapted specifically for SPHEREx data, which involves hundreds of filters and demands high-throughput processing of large datasets.

The code performs $\chi^2$ minimization over a predefined grid of models using 160 templates, including those from the Brown and COSMOS libraries. The model grid spans $E(B-V)$ values from 0 to 1 in steps of $\Delta E(B-V) = 0.1$ for three dust extinction laws \citep{allen76, prevot84, calzetti00}, and covers redshifts from $z = 0$ to 3 in steps of $\Delta z = 0.002$. 
The grid size is determined after careful optimization. 
Although this redshift range limits the applicability of the models to sources at $z \leq 3$, we do not extend the grid to higher redshifts because galaxies at $z > 3$ are expected to be faint and relatively rare in SPHEREx data. Flat priors are assumed over the parameter space and template set. 

The code has thus far been tested only on simulated datasets. Priors and the template error function will be finalized and validated once SPHEREx data becomes available. Note that the expected value of the redshift probability distribution function (PDF) is adopted as the point estimate. Negative flux measurements are excluded from the fitting procedure.

\subsection{EAZY} \label{sec:method_eazyc}
EAZY \citep{brammer08} is a photo-$z$ code written in the C language. 
This software is widely used to measure photo-$z$, along with its Python version that will be described below.
This code estimates photo-$z$ based on $\chi^2$ minimization. 
We can additionally impose a magnitude-redshift prior to exploit the trend of galaxies to be further with fainter magnitudes.

The size of the redshift grid is important for the precision of the photo-$z$ measurements.
When the grid is too fine, the calculation time becomes significantly longer. 
However, the size of the grid does not have a significant influence on the results when the S/N of the data is low. 
To balance between the precision of the measurements and the calculation time, we used magnitude-dependent redshift grids.
We use grid size of \texttt{Z\_STEP} = 0.01 for $i > 19$ targets and \texttt{Z\_STEP} = 0.001 for $i \leq 19$ targets.

As noted above, we use the Brown + COSMOS templates from \cite{feder24} for SED fittings.
However, among the Brown + COSMOS templates, the COSMOS templates do not have information for wavelengths shorter than 900\r{A}, which overlap with the 7DT wavelengths from $z \geq 3.44$. 
The majority of the objects in the catalogs have lower redshifts, but some objects may be influenced by the incomplete short wavelength coverage of the templates in photo-$z$ estimation.
Therefore, we did not use the COSMOS template, leaving 129 Brown templates in the photo-$z$ estimation.
We adjust the corresponding configurations, such as the wavelength file, to the templates.
While EAZY and eazy-py support multi-template fitting --- allowing the observed fluxes to be modeled as a linear combination of templates --- this option substantially increases the computational cost.
Given that the template set captures the main features of the SEDs observed in our sample, we opted for single-template fitting to reduce runtime without significant loss of accuracy.

The template error function is assumed to be constant at 5\,\% for each template in all wavelengths (c.f. \cite{rudnick01}).
When we use a non-zero template error function, the accuracy of the estimated photo-$z$ improves as it smooths the resulting probability distribution of the photo-$z$ posterior and increases the influence of the prior.
This prevents the photo-$z$ estimation from being a less plausible but global minimum in the posterior distribution.
We also test a photo-$z$ performance using a template error function following \cite{brammer08} and found that the results remain similar.
We therefore decided to adopt constant error functions in this study, as the error function constructed in this experiment is template- and sample-dependent and thus is not suitable for application to future real datasets, while also adding unnecessary complexity to the comparison.
We use \texttt{z\_m2} as our point estimate from the resulting probability distribution function (PDF), which is calculated after marginalizing the magnitude-redshift prior.

We construct a prior from the redshift-magnitude distribution of the galaxies for a certain band in the dataset, as the SPHEREx and 7DS have unique sets of bands.
We use $m625$ magnitude in the prior for 7DS and SPHEREx + 7DS datasets, as this band has the highest expected S/N among the SPHEREx and 7DS bands.
When we measure photo-$z$ values only with SPHEREx data, we use the 14th channel of detector 3 in SPHEREx, as this channel has the highest expected S/N.

We construct the prior by fitting the magnitude-redshift distribution of the COSMOS dataset with the functional form from Equation (3) of \cite{brammer08}.
The COSMOS2020 \citep{weaver22} data have an $i$-magnitude depth of $\lesssim 27$ magnitude. 
This is much deeper than $i \sim 23$ magnitude, which is the magnitude of the faintest galaxy that we present a photo-$z$ in this study.
Most of the bands used in COSMOS2020 have depth deeper than 23 $i$-magnitude, which makes the effect from the number of data points used in SED construction of \cite{feder24} minimal for galaxies with $i \lesssim 23$.
Therefore, the COSMOS2020 photometric catalog and its corresponding mock SEDs built in \cite{feder24} are complete down to $i \sim 23$ magnitude, and thus we use this sample to make a prior representative of the distribution of the galaxies within a specific volume in the real Universe.
The resulting shape of the prior is very similar to the one illustrated in Figure 4 in \cite{brammer08}.

\subsection{eazy-py} \label{sec:method_eazypy}
The photo-z estimator, eazy-py \citep{brammer21} is a Python implementation of EAZY, designed to reproduce the results of the original C version while offering improved flexibility in Python-based workflows. 
This software supports most of the functionalities provided by EAZY, including multi-template fitting, the use of template error functions, and redshift priors. 
Owing to its reliability and integration within the broader Python ecosystem, eazy-py has become one of the most widely adopted photo-$z$ estimation tools in the current era of the James Webb Space Telescope and ongoing medium- and narrow-band imaging surveys \citep[e.g., ][]{lee24}.

Although the EAZY and eazy-py share many properties, we used different redshift grids, template sets, template error, and priors, as we tuned the settings independently.
We will discuss the difference in the results in Section \ref{sec:dis_comparison}.

We estimate photo-$z$ using the 160 Brown+COSMOS templates with single-template fitting. 
We use single-template fitting for the same reason as in the EAZY C version (see Section \ref{sec:method_eazyc} for more details).

The redshift grid is defined with a step size of \texttt{Z\_STEP}=0.01, spanning the range 0.002 to 5.8, chosen to encompass the redshift distribution of the sample. A constant template error of 0.01 is applied across all wavelengths. We also scale the flux errors by a factor of 1.5 to improve the calibration of the redshift PDF, $p(z)$, in separate validation tests using our deep dataset following the methods of \citet{wittman16} and \citet{laur22_topz}.
Further discussion of this point is provided in Section \ref{sec:dis_confidence}

As for redshift priors, we adopt the built-in $K_{\rm s}$-band prior (``\texttt{prior\_K\_TAO}''), derived from the Theoretical Astrophysical Observatory (TAO) lightcone simulation \citep{bernyk16_tao}. For details on the prior-generation methodology, see \citet[Section 2.5]{brammer08}. This prior is linked to the 14th channel of SPHEREx band 3, which has a pivot wavelength of 2.2 $\mu$m, corresponding to the $K_{\rm s}$ band. For the 7DS-only sample, we instead apply the built-in R-band prior (``\texttt{prior\_R\_zmax7}''), derived from the luminosity function of the semi-analytic model in \citet{delucia07}, and link it to the 7DT/\textit{m}650 band. 

Optimization is performed using the default non-negative least squares algorithm implemented in eazy-py \citep{lawson74, scipy}, and all negative flux values are excluded from the fitting. The final photo-$z$ point estimate is taken to be the maximum a posteriori (MAP) redshift (\texttt{zml}), which is also the default output of eazy-py.

\subsection{LePhare} \label{sec:method_lephare}
The LePhare \citep{arnouts99} is a photometric redshift estimation tool that utilizes the spectral energy distribution (SED) template fitting method. In addition to redshifts, it can derive physical properties such as stellar mass, star formation rate, and dust content. As one of the longest-standing tools, LePhare has been widely adopted as a standard photometric redshift estimator in large surveys such as COSMOS \citep{weaver22} and Canada-France-Hawaii Telescope Legacy Survey CFHTLS (CFHTLS; \citealp{ilbert06, coupon09}).

Among the parameters in LePhare, we set Z\_STEP, which determines the step size of the redshift grid used for SED fitting, to 0.005. This value is chosen to balance accuracy and computational efficiency. 
We also test Z\_STEP = 0.002, but the redshift performance does not significantly improve even for bright GAMA galaxies.
The best-fit redshifts are determined from the PDF computed with the $\chi^2$ minimization on a redshift grid, with parabolic interpolation applied to refine the grid beyond the resolution set by Z\_STEP. 
No specific prior is applied, and neither dust extinction nor dust emission is considered, as these effects are already incorporated into the SED templates. 
Non-detections, which correspond to upper limits, can be handled in two ways: either excluded from the fitting (with –99 values), or used as upper-limit constraints (by setting error = –1 and assigning fluxes at the 3$\sigma$ or 5$\sigma$ level). 
In the SED fitting with LePhare, we discard non-detections exhibiting negative fluxes.
We use Z\_BEST as the point estimate for the redshifts.

\subsection{Deep Neural Network (DNN)} \label{sec:method_dnn} 
DNN is a machine learning model composed of multiple hidden layers, which excels at progressively learning complex and abstract features from data as it passes through each layer.
Although in modern usage DNN typically means deeper models, we use the term ``Deep'' as we utilize multiple hidden layers, following the more conventional definition.
While template-fitting methods rely on a limited set of predefined galaxy spectral templates, DNNs can learn these intricate nonlinear relationships directly from large-scale observational data. 

The network employed in this study consists of an input layer, an output layer, and five hidden layers. The hidden layers are generated with the number of neurons changing as follows: $128 \rightarrow 256 \rightarrow 512 \rightarrow 256 \rightarrow 128$. 
This structure encourages the effective compression and expansion of information. 
For all hidden layers, we used the Rectified Linear Unit (ReLU) as the activation function.
To prevent overfitting, a dropout rate of 20\,\% was applied immediately after each of the first four hidden layers to enhance the model's generalization performance. 
The final output layer consists of a single neuron that yields the predicted redshift value.
We also test deeper layers with 8 hidden layers, but did not find improvements in the performance for the data.

The Mean Squared Error (MSE), a widely used metric in regression tasks, aims to minimize the square sum of the residuals ($z_{\text{phot}} - z_{\text{true}}$). 
The MSE tends to be dominated by the variance, while a small but persistent systematic bias contributes weakly to the total loss. 
To mitigate this issue, we introduce a `bias-corrected loss function' that has an additional term that penalises the magnitude of the bias more heavily than the standard MSE to reduce the bias of the resulting photo-$z$s.

\begin{equation}
L_{\text{total}} = \underbrace{\frac{1}{N} \sum_{i=1}^{N} ( \hat{z}_{i} - z_{i} )^{2}}_{L_{\text{MSE}}} + \lambda \underbrace{\left| \frac{1}{N} \sum_{i=1}^{N} (\hat{z}_{i} - z_{i}) \right|}_{\text{Bias Term}}
\end{equation}
Here, $\lambda$ is a hyperparameter that controls the weight of the bias term, which is set to be 0.1. The second term (Bias Term) directly computes the absolute value of the mean residual—the bias—and incorporates it into the loss. 
The bias-corrected loss function penalizes more steeply as a function of bias compared to the ordinary MSE.
We tested both the bias-corrected loss function and the standard MSE, finding that the former yields superior performance in terms of bias, scatter, and outlier fraction.
For example, the bias-corrected loss function reduces the bias by about 85\,\% for the combined SPHEREx + 7DS GAMA dataset with wide depth.
The model with a bias-corrected loss function shows significantly higher performance in reducing the bias between $z_{phot}$ and $z_{true}$ in redshift estimations.
Consequently, we adopt the bias-corrected loss function in the model.

We use `standardized color indices' as the input features for the model.
Using the input features, the DNN model infers the output photo-$z$ values.
The `standardized color indices' are calculated through the following procedures to improve the quality of the mock catalog and the performance of the model.

First, we replace any missing (NaN) values with the interpolated values from the adjacent bands, if they are not NaN values.
When the interpolation is not available, we use 1$\sigma$ limiting flux values.
We also substitute the negative flux values with the 1$\sigma$ limiting flux values, too.
Second, we inject Gaussian noise into the training set using the fiducial error of each data point. 
This procedure acts as a regularization technique to prevent model overfitting and improve generalization.

After then, we calculate the color index using the following equation:
    $$
    c_i = 2.5 \log_{10} \left( \frac{f_i}{f_{\text{central}}} \right)
    $$
Here, $c_i$ is the color index for band $i$. A small constant ($10^{-8}$) was added to both the numerator and the denominator to avoid numerical issues with zero fluxes.
$f_{i}$ is the flux in band $i$ and $f_{\text{central}}$ is the flux in the band located at the midpoint of the wavelength range, such as the 51st band for SPHEREx.
Finally, we standardize the data using the \texttt{StandardScalar} from scikit-learn \citep{pedregosa11} to make each feature have a zero mean and unit variance for enhancing the training efficiency of the model.
In summary, the final input features for our model are the `standardized color indices'.

For model training and performance evaluation, the entire dataset is divided into two independent sets based on the `$\texttt{FLAG\_ML}$' flag in the source catalog. No separate test set is used in this study; modeling is performed using only training and test sets.
We use the training set to train the weights and biases of the model.
The model learns the underlying relationships between the color indices and the true redshift from this data.
The test set ($\texttt{FLAG\_ML=1}$) is not used in the training process. 

The model is trained using the PyTorch Lightning framework to ensure reproducibility and automation. Using the training set, the model is optimized with the Adam optimizer, with an initial learning rate of $1 \times 10^{-4}$ and a mini-batch size of 512. The training was scheduled for a maximum of 10,000 epochs.
To prevent overfitting and to terminate the training efficiently, we employ an early-stopping strategy. After each epoch, the loss is evaluated on the validation set. If this validation loss does not show improvement for 2,000 consecutive epochs, the training process is automatically terminated. This strategy ensures that training stops at the point where the model's generalization performance is no longer improving, thereby preventing overfitting and securing a final model with optimal performance.

The DNN model demonstrates excellent performance for bright objects with high signal-to-noise ratios, such as those in the GAMA data. 
However, we do not apply it to datasets containing a large number of faint objects, like our simulated COSMOS sample. 
In the simulated COSMOS sample, the low signal-to-noise ratios of faint galaxies occasionally result in negative flux values across multiple bands, which arise from random sampling when we generate the mock catalog.
A high prevalence of these non-physical values can hinder the model’s ability to learn the true underlying relationship between color and redshift. 
Although our preprocessing pipeline replaces negative fluxes with forced positive values, this imputation is not a fundamental solution and risks introducing distorted features. 
For these reasons, we concluded that applying the DNN model to the COSMOS data would be unlikely to yield reliable results and therefore excluded it from this part of the analysis.

\subsection{Hierarchical Random Forests (HRF)} \label{sec:method_rf}
The random forest is a machine learning model based on the decision tree algorithm. We build a  model that combines the random forests in hierarchical order (hierarchical random forest, HRF, \citealp{kim25a}). Here, we describe the structure of our model, including the decision tree and random forest on which our model is based.

The decision tree is the building block of the random forest. The decision tree performs classification or regression by recursively splitting the input space (i.e., flux and flux uncertainties for our case) into sub-spaces in which the outputs are similar, based on the training set \citep{James2021}. Specifically, for regression, the decision tree aims to minimize the variance of outputs, defined as the residual sum of squares (RSS):
\begin{equation}
    \text{RSS}(J) = \sum^J_{j=1}\sum_{i\in S_j}(z_{ij}-\overline{z}_j)^2,
    \label{RSS}
\end{equation}
where $z_{ij}$ is the output (redshifts for our case) of $i-$th element in the $j$-th sub-space and $\overline{z}_{j}$ is the mean of $z_{i}$ in $j$-th sub-spaces. The decision tree first calculates the RSS of the entire training set ($J$=1). The decision tree then determines the split of input space ($J=2$) that maximizes the decrease in RSS from $J=1$. By repeating the splits with increasing ($J$), the decision tree determines the splits that maximize the decrease in RSS for a certain depth ($J$). In other words, it finds groupings of inputs that share similar outputs.

For classification, the decision tree works similarly to regression, except that it aims to minimize the heterogeneity of outputs, defined as Gini impurity:
\begin{equation}
    \text{Gini}(J) = \sum^J_{j=1}\sum_{k\in K}(1-p_{k,j}^2).
    \label{RSS}
\end{equation}
Here, we assume that the output is categorical with $K$ values and $p_{k,j}$ is the frequency of training data with output $k$ in the $j$-th subspace. Because $\sum_{k\in K}(1-p_{k,j}^2)=0$ when $p_{k',j}=1$ and $p_{k\neq k',j}=0$ for a certain $k'\in K$, minimizing Gini impurity indicates grouping training data with similar outputs. 

The random forest is an ensemble of many decision trees \citep{Breiman2001, James2021}. It bootstraps the training set into $N_{\rm bootstrap}$ sub-sets and trains the decision tree for each sub-set. At each splitting step, decision trees randomly select features used for prediction at each split. The random selection of the features at each split prevents the random forest from overfitting.
When it makes a prediction, the random forest returns the mean output of all the decision trees as its final output. 
In our random forest model, the number of bootstrapped training sets ($N_{\rm boots}$) is 30, and the number of selected features ($N_{\rm feature}$) is one-third of the total number of input features. We test the performance of the random forest models on our sample by changing $N_{\rm boots}$ and $N_{\rm feature}$. The performance was almost similar regardless of $N_{\rm boots}$ and $N_{\rm feature}$.

Ensuring a uniform distribution of the training set is crucial for achieving unbiased predictions across the entire output range for the random forest. If the training set is skewed to a specific redshift range, most splits of the random forest occur within that range to minimize RSS. Consequently, the random forest sparsely partitions redshift ranges with fewer training samples. Because the random forest performs regression by averaging the outputs of the training outputs, the coarse partitioning results in systematic overestimation or underestimation. Thus, uniformizing the training set is necessary to produce homogeneous predictions across the full output range.

We develop the HRF that estimates the redshifts using the uniformly distributed training sample by hierarchically combining the random forest models. The HRF consists of the classification phase that uniformizes the distribution of training samples and the regression phase that infers the redshifts in the uniform training sample from the previous phase. 

In the classification phase, our model recursively performs binary classifications to predict the redshift range to which an object belongs. At each binary classification stage, we select a threshold above and below which the number of training samples is similar. We then train a classification random forest model to classify whether the redshift corresponding to a given photometric input falls below or above the selected threshold. 
We iterate the binary classifications up to the fourth stage, where the resulting redshift bins have a uniform distribution.

In the regression phase, the HRF estimates the redshift using a regression random forest. We train the regression random forest on each of the redshift bins created during the final classification stage. Because each bin contains a uniform distribution of training samples, the regression models trained on these bins are free from the biases resulting from the skewed distribution.

The input of our model is photometry, including flux and flux uncertainties, and the output is the photometric redshift.
We replace the negative flux values with a dummy value of 1,000 to maintain the dimensionality of the input data \citep{cohen75}.
The dummy value, 1,000, significantly differs from the general flux values, locating the input with the dummy value far from other inputs in the input space. The random forest then learns not to use this dummy value in the prediction \citep{kim25a}.

\section{Results} \label{sec:result}
In the following sections, we use three statistics to quantify the performance of photo-$z$ measurements. 
We define $\Delta z = z_{phot} - z_{true}$ in the equations below.
We do not apply sigma-clipping while calculating these statistics.
The errors of the statistics were calculated using a bootstrap method by randomly sampling 1,000 subsamples.

\begin{itemize}
  \item Normalized Median Absolute Deviation (NMAD), $\sigma$, represents the scatter of photo-$z$ measurements for a given sample. 
  This metric is widely used to quantify the scatter of the photo-$z$ values around the true redshifts (e.g., \citealp{dahlen13, laur22_topz, feder24, ko25}) and it is more stable to outliers compared to the standard deviation of $\Delta z / (1 + z_{true})$.
  \begin{equation}
    \sigma = 1.4826\times \text{median}(|\Delta z - \text{median}(\Delta z)|/(1+z_{true}))
    \label{eq:sigma}
   \end{equation}
   
  \item Bias, $b$, shows the systematic offset of photo-$z$ values from true redshifts for a given sample of galaxies.
  \begin{equation}
    b = \text{mean}((\Delta z)/(1 + z_{true}))
    \label{eq:bias}
   \end{equation}
 
  \item Catastrophic failure, $\eta$, measures the fraction of galaxies with severely deviated photo-$z$ values from their true redshifts.
  Following previous research \citep{tanaka18, ko25}, we define $\eta$ to be the outlier with its fractional deviation larger than 10\,\%.
  \begin{equation}
  \eta = \text{fraction}(|\Delta z|/(1 + z_{true}) > 0.1)
  \label{eq:eta}
  \end{equation}
  
\end{itemize}

Although template-fitting codes often yield different point estimates for the final redshift, in this study, we adopt the most commonly recommended estimates, as we aim to demonstrate the potential of the novel data set when used with widely employed codes.

\begin{figure*}
\includegraphics[width=\textwidth]{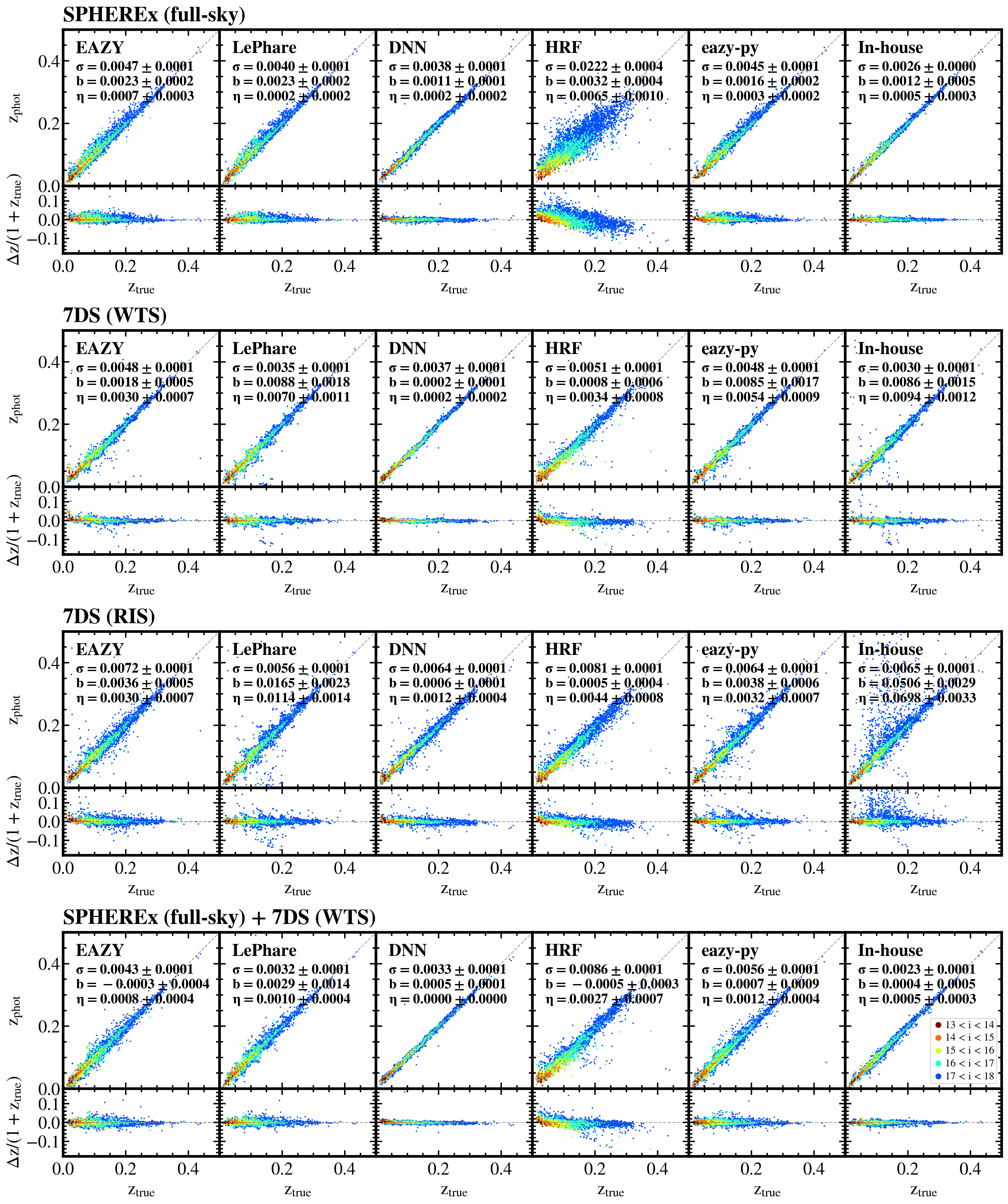}
\caption{The measured photo-$z$ values as a function of true redshifts. GAMA galaxies with $13<i<18$ ($N=$ 5,961) in the wide dataset are shown in this figure. The colors show the magnitudes of the galaxies, binned by 1-magnitude intervals. From the first row, results from the SPHEREx full-sky data, 7DS WTS data, 7DS RIS data, and combined SPHEREx full-sky + 7DS WTS data are plotted. Each plot has a subplot showing $\Delta z/(1+z_{\rm true})$ as a function of $z_{\rm true}$. The statistics for measuring photo-z evaluation performances, NMAD ($\sigma$), bias ($b$), catastrophic failure ($\eta$, 10\,\% outlier rate), and their corresponding errors, are noted in each plot.
 \label{fig:zz_1318}}
\end{figure*}

\begin{figure*}
\includegraphics[width=\textwidth]{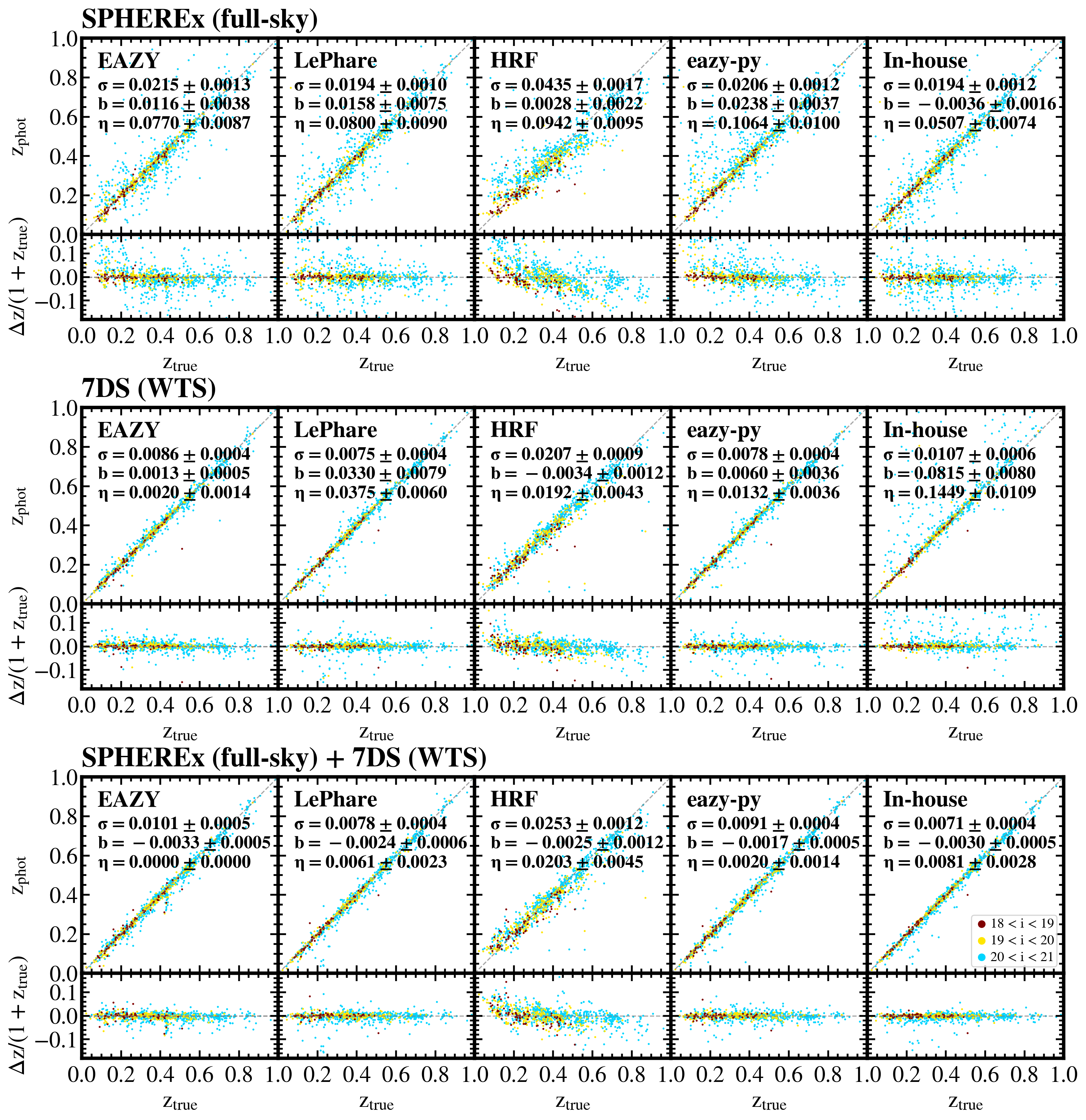}
\caption{Same as Figure \ref{fig:zz_1318}, but for COSMOS galaxies with $18<i<21$ ($N=$ 990) in the wide dataset. Note that the redshift and magnitude ranges of the sample are different from Figure \ref{fig:zz_1318}.
 \label{fig:zz_1823}}
\end{figure*}

\begin{table*}
\caption{Statistics of COSMOS galaxies in the wide ($N=34,531$) and deep dataset ($N=219$) with $18 < i < 23$.}
\label{tab:stats_wide_deep}
\centering
\begin{tabular}{cc|r|r|r|r}
\hline
\multirow{2}{*}{Data} & \multirow{2}{*}{Stat.} & \multicolumn{2}{c}{EAZY}& \multicolumn{2}{|c}{LePhare}\\
\cline{3-6}
\multicolumn{1}{c}{} & \multicolumn{1}{c}{} & \multicolumn{1}{|c}{Wide} & \multicolumn{1}{|c}{Deep} & \multicolumn{1}{|c}{Wide} & \multicolumn{1}{|c}{Deep}
\\
\hline
{       }      & $\sigma$& 0.2737$\pm$0.0029 & 0.0152$\pm$0.0017 & 0.3570$\pm$0.0066 &  0.0129$\pm$0.0017 \\ 
{SPHEREx}      & $b$     & 0.3004$\pm$0.0026 & -0.0078$\pm$0.0032 & 0.6348$\pm$0.0049 &  -0.0037$\pm$0.0036 \\
{       }      & $\eta$  & 0.6216$\pm$0.0025 & 0.0411$\pm$0.0129 & 0.6113$\pm$0.0026 &  0.0457$\pm$0.0139 \\
\hline
{       }      & $\sigma$& 0.0720$\pm$0.0009 & 0.0159$\pm$0.0018 & 0.1269$\pm$0.0022 &  0.0137$\pm$0.0021 \\ 
{7DS}          & $b$     & 0.0514$\pm$0.0009 & 0.0150$\pm$0.0043 & 0.2920$\pm$0.0036 &  0.0430$\pm$0.0178 \\
{       }      & $\eta$  & 0.3369$\pm$0.0027 & 0.0868$\pm$0.0190 & 0.4751$\pm$0.0028 &  0.1142$\pm$0.0213 \\
\hline
{       }      & $\sigma$& 0.0510$\pm$0.0005 & 0.0128$\pm$0.0013 & 0.0438$\pm$0.0006 & 0.0103$\pm$0.0014 \\ 
{SPHEREx + 7DS}& $b$     & 0.0189$\pm$0.0008 & -0.0084$\pm$0.0022 & 0.0731$\pm$0.0023 & -0.0040$\pm$0.0019 \\
{       }      & $\eta$  & 0.2142$\pm$0.0022 & 0.0183$\pm$0.0093 & 0.2505$\pm$0.0022 & 0.0046$\pm$0.0047 \\
\hline
\hline
\multirow{2}{*}{Data} & \multirow{2}{*}{Stat.} & \multicolumn{2}{|c}{eazy-py} & \multicolumn{2}{|c}{In-house}\\
\cline{3-6}
\multicolumn{1}{c}{} & \multicolumn{1}{c}{} & \multicolumn{1}{|c}{Wide} & \multicolumn{1}{|c}{Deep} & \multicolumn{1}{|c}{Wide} & \multicolumn{1}{|c}{Deep}
\\
\hline
{       }      & $\sigma$& 0.1527$\pm$0.0020 & 0.0106$\pm$0.0015 & 0.1337$\pm$0.0014 &  0.0109$\pm$0.0019 \\ 
{SPHEREx}      & $b$     & 0.2366$\pm$0.0033 & -0.0028$\pm$0.0030 & 0.0901$\pm$0.0014 &  0.0026$\pm$0.0028 \\
{       }      & $\eta$  & 0.5025$\pm$0.0027 & 0.0365$\pm$0.0128 & 0.4672$\pm$0.0027 &  0.0320$\pm$0.0119 \\
\hline
{       }      & $\sigma$& 0.0573$\pm$0.0008 & 0.0127$\pm$0.0017 & 0.2677$\pm$0.0022 &  0.0274$\pm$0.0058 \\ 
{7DS}          & $b$     & -0.0305$\pm$0.0009 & -0.0123$\pm$0.0066 & 0.3887$\pm$0.0023 &  0.0731$\pm$0.0107 \\
{       }      & $\eta$  & 0.2949$\pm$0.0024 & 0.0639$\pm$0.0165 & 0.7561$\pm$0.0023 &  0.2557$\pm$0.0296 \\
\hline
{       }      & $\sigma$& 0.0393$\pm$0.0004 & 0.0085$\pm$0.0010 & 0.0589$\pm$0.0007 & 0.0087$\pm$0.0007 \\ 
{SPHEREx + 7DS}& $b$     & 0.0373$\pm$0.0016 & -0.0024$\pm$0.0020 & 0.0009$\pm$0.0009 & -0.0031$\pm$0.0014 \\
{       }      & $\eta$  & 0.1939$\pm$0.0021 & 0.0046$\pm$0.0045 & 0.2912$\pm$0.0024 & 0.0046$\pm$0.0046 \\
\hline
\end{tabular}
\tablefoot{We do not include the statistics from the ML-based methods (RF and DNN) in this table, as the number of test samples in this magnitude range ($N=51$) is too small to derive reliable statistics for comparison between the methods.}
\end{table*}

\subsection{Photo-$z$ Estimation Performance} \label{sec:res_photoz}
We present the plots showing the distribution of the measured photo-$z$ values compared to their true redshifts in Figure \ref{fig:zz_1318} and \ref{fig:zz_1823}. 
The figures show the results for GAMA and COSMOS galaxies in the wide dataset, respectively.
Table \ref{tab:stats_wide_deep} presents the statistical results for galaxies in the range $18 < i < 23$ from the template-fitting methods, reported separately for the wide and deep datasets. 

Figure \ref{fig:zz_1318} shows the distribution of measured photo-$z$ values from different methods and different survey combinations for the bright ($13 < i < 18$) GAMA galaxies as a function of true redshifts.
To enable robust comparison between the template fitting methods and ML-based methods, we use only the galaxies in the test sample for the figures.
In Figure \ref{fig:zz_1318}, the distribution of photo-$z$ values shows strong one-to-one correlations with the true redshifts of the mock SEDs, although the scatter of the resulting photo-$z$ values is different for different combinations of methods and data.
The brighter galaxies show a tighter one-to-one correlation compared to the fainter ones in all the plotted combinations of surveys and methods.
The values of statistics are $\sigma \lesssim 0.005$, $b \lesssim 0.005$, and $\eta \lesssim 0.005$ across all the combinations of methods and surveys presented in Figure \ref{fig:zz_1318}, with the best case reaching down to 0.23\,\% accuracy.

Figure \ref{fig:zz_1823} shows the measured photo-$z$ values as a function of true redshifts for the COSMOS galaxies with $18<i<21$ in the wide dataset.
We use only the galaxies in the test sample.
As in Figure \ref{fig:zz_1318}, we plot the results for each combination of methods and surveys separately.
The combined SPHEREx and 7DS data show $\sigma \sim 0.015$, $|b| \sim 0.005$, and $\eta \sim 0.003$ for all the methods used, with the best case showing 0.71\,\% accuracy.
We do not have measurements from the DNN method of COSMOS galaxies, as the DNN method suffers from many non-detections in the faint galaxies (see Section \ref{sec:method_dnn}).

Table \ref{tab:stats_wide_deep} demonstrates that the overall results from the deep dataset are better than those from the wide dataset.
The $i$-band magnitude limit of 23 is comparable to the median depth of the deep sample.
Therefore, Table \ref{tab:stats_wide_deep} shows roughly similar statistics to the galaxies with $18 < i < 21$ in the wide dataset, which is summarized in Figure \ref{fig:zz_1823}.
In other words, the deep dataset shows comparable photo-$z$ performance to the wide dataset, even after including the galaxies that are fainter by two magnitudes.

When we exclude outliers for estimating the photo-$z$ statistics, the $\sigma$ for the COSMOS SPHEREx + 7DS sample with $18<i<23$ improves by about 30\,\%, for example.
This suggests that we can improve photo-$z$ performance significantly if we can identify outliers in advance.
The development of schemes for identifying outliers is being considered.

\subsection{Magnitude Dependence of Photo-$z$ Performance}
\begin{figure*}
\includegraphics[width=0.95\textwidth]{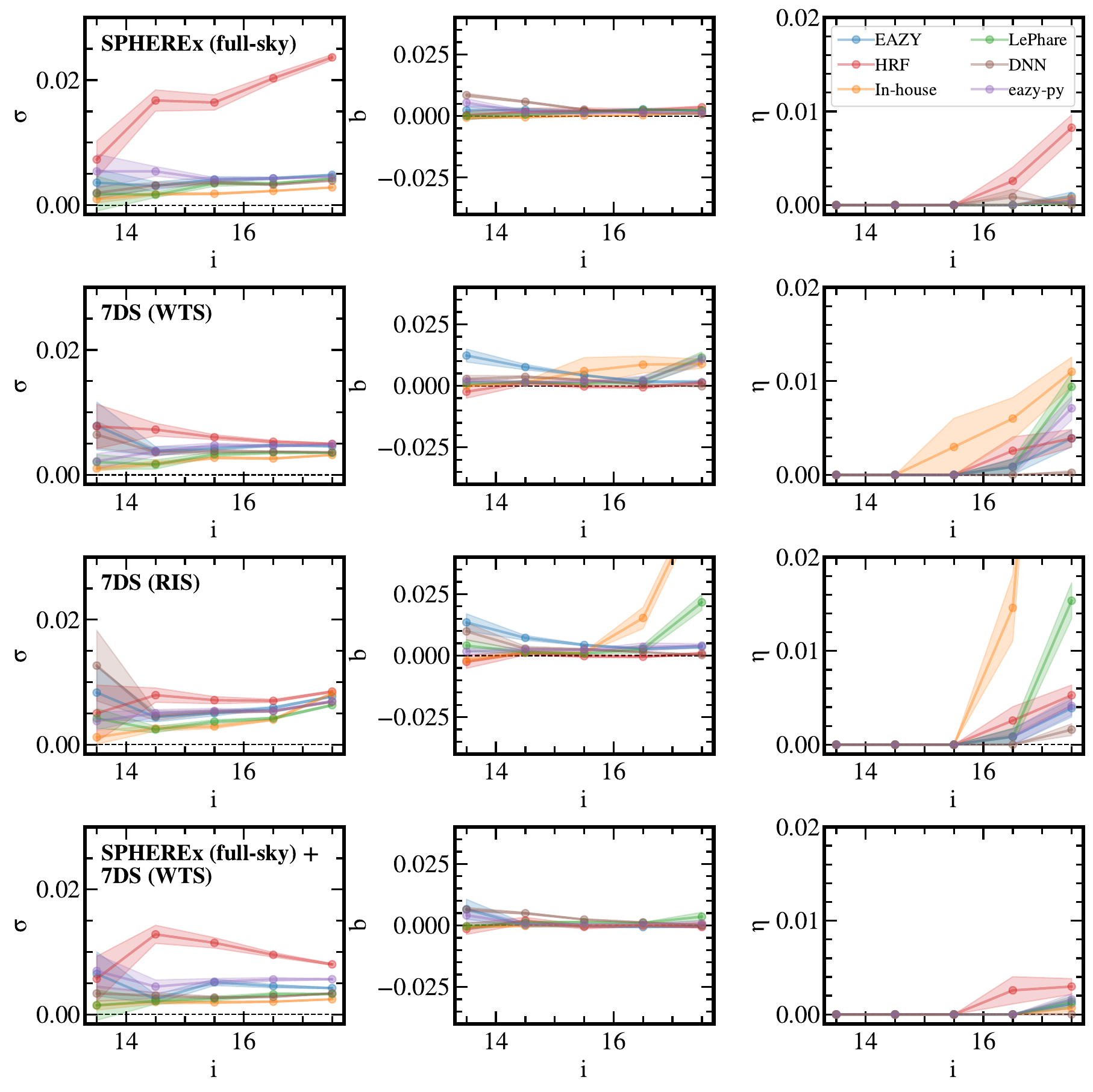}
\caption{The statistics of photo-$z$ values as a function of $i$-magnitudes.
The GAMA galaxies within the wide dataset are used for calculating statistics.
From the first row in the figure, we plot the statistics from the SPHEREx full-sky, 7DS WTS, 7DS RIS, and combined SPHEREx full-sky + 7DS WTS data from the first row to the bottom. 
The statistics from different methods are plotted with different colors, which are noted in the legend.
The shaded region interpolates the confidence intervals of the bins.
 \label{fig:stats_gama}}
\end{figure*}

\begin{figure*}
\includegraphics[width=0.95\textwidth]{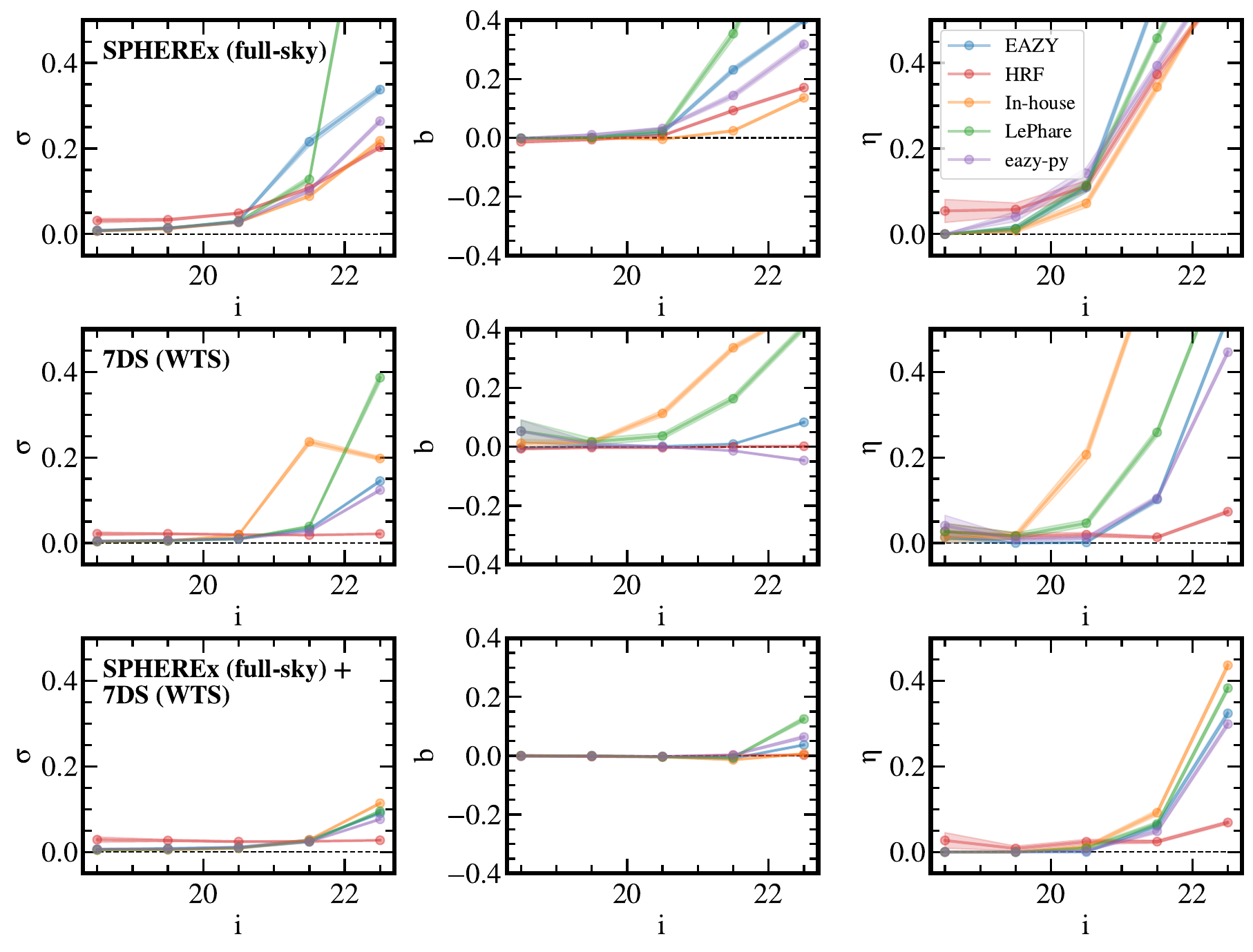}
\caption{Similar to  Figure \ref{fig:stats_gama}, but for the fainter COSMOS galaxies within the wide dataset.
We plot the results from the SPHEREx full-sky, 7DS WTS, and combined SPHEREx full-sky + 7DS WTS data from the first row. 
Note that the magnitude ranges and y limits are different from those of Figure \ref{fig:stats_gama}.
 \label{fig:stats_cos}}
\end{figure*}

As we noted in the previous section, the resulting performance of photo-$z$ is strongly dependent on the magnitudes of the galaxies. 
In this regard, we plot the statistics as a function of $i$-band magnitudes in Figures \ref{fig:stats_gama} and \ref{fig:stats_cos} using the wide dataset for GAMA ($13 < i< 18$) and COSMOS ($18 < i < 23$) galaxies, respectively.
We use only test samples for these figures to enable comparison between the ML-based methods and template fitting methods, as we did in Figures \ref{fig:zz_1318} and \ref{fig:zz_1823}.
We plot $\sigma$, $b$, and $\eta$ in different columns. 
We show the results of different survey data in the different rows of the figure.
The methods are differentiated by the colors. 

Figure \ref{fig:stats_gama} shows the dependence of the photo-$z$ performance as a function of $i$ magnitudes for the GAMA galaxies in the wide dataset.
As they already provide sufficient signal to the SPHEREx full-sky and 7DS WTS surveys even for the faintest target plotted ($i\approx18$), they show more precise and robust photo-$z$ estimates compared to the fainter COSMOS galaxies. 
Also, the trend between the statistics and $i$ magnitudes is mostly flat, except for some cases like the $\sigma$ of SPHEREx when using the RF method or the $\eta$ values from SPHEREx in-house code for 7DS-only data.
The results from the RIS data also show comparable statistics to the deeper WTS data.

In Figure \ref{fig:stats_cos}, we plot the relation between the photo-$z$ statistics and $i$ magnitudes for the COSMOS galaxies in the wide dataset.
We do not measure the photo-$z$ for the COSMOS sample with the DNN method because of the difficulty of handling many non-detections (see Section \ref{sec:method_dnn}).
The relation between $i$ magnitude and the statistics in Figure \ref{fig:stats_cos} clearly shows the dependence of photo-$z$ estimation performance on the S/N of the data.
When only the SPHEREx data is used for photo-$z$ estimation, the statistics become worse for the galaxies with $i>21$, as it is below the detection threshold (See Figure \ref{fig:depth}).
The 7DS data shows a comparable level of statistics to the SPHEREx data, although the statistics from different methods show larger scatter.
For SPHEREx + 7DS data, the $b$, $\sigma$, and $\eta$ are stable for the galaxies with $i < 22$, exhibiting better performance compared to either 7DS- or SPHEREx-only data.

\subsection{Redshift Dependence of Photo-$z$ Performance}

\begin{figure*}
\includegraphics[width=0.95\textwidth]{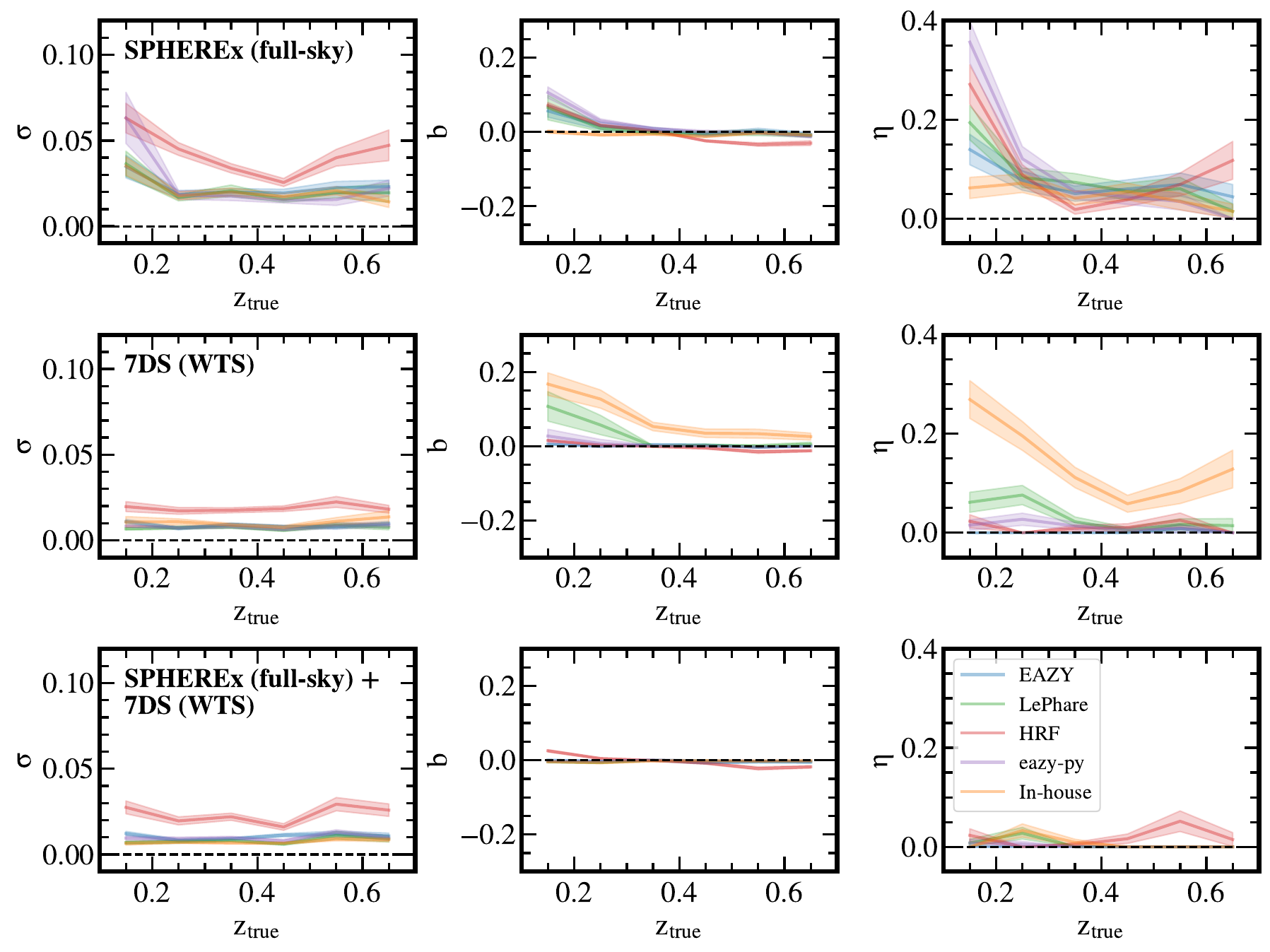}
\caption{The statistics of photo-$z$ values as a function of true redshifts ($z_{\rm true}$).
COSMOS galaxies with $18 < i < 21$ in the wide dataset are used to calculate the statistics.
The number of galaxies within each $z_{\rm true}$ bin is higher than 70.
From the first row in the figure, we plot the statistics from SPHEREx, 7DS, and SPHEREx + 7DS data.
The shaded region interpolates the confidence intervals of the bins.}
\label{fig:z_stat}
\end{figure*}

We plot the dependence of the measured statistics on the true redshifts ($z_{\rm true}$) in Figure \ref{fig:z_stat}.
In this figure, we use the results of COSMOS galaxies in the wide dataset with $18 < i < 21$.
We calculate statistics of the galaxies within each bin of the true redshift ($\Delta z_{\rm true} = 0.1$).
We calculate the statistics only for bins containing more than 70 galaxies to ensure robust estimation. 
We use the galaxies within the test sample for fair comparison between the ML-based and template fitting methods.

The results in the last row from the combined SPHEREx + 7DS data show the lowest level of $\sigma$, $b$, and $\eta$ across all true redshifts plotted in the figure.
All methods plotted in the figure exhibit the improved statistics for the combined dataset compared to either the SPHEREx or 7DS dataset.
We will further discuss the synergy between the SPHEREx and 7DS in Section \ref{sec:dis_synergy}.

\section{Discussion} \label{sec:discussion}

\subsection{Photo-$z$ with SPHEREx and 7DS} \label{sec:dis_photoz}
The SPHEREx and 7DS surveys densely sample galaxy SEDs from optical to near-infrared wavelengths without spectral gaps, covering hundreds of millions of galaxies across wide sky areas without target pre-selection.
In this study, we forecast the precision and accuracy of photo-$z$ measurements from these surveys using simulated datasets.

The continuous sampling of SEDs significantly enhances the highest achievable precision of photo-$z$ values, particularly for bright galaxies with enough S/N for relatively narrow bandwidths.
Therefore, our results show $\sigma$, $b$, and $\eta$ values lower than 0.01 for all the survey combinations and most of the methods when we measure photo-$z$ values of GAMA galaxies ($13 < i <18$, see Figure \ref{fig:zz_1318}) within the wide dataset (SPHEREx full-sky + 7DS WTS).

Furthermore, the SPHEREx full-sky survey and RIS show comparable photo-$z$ measuring performance to deeper WTS or combined SPHEREx + WTS, despite their relatively shallower depths (see Figure \ref{fig:stats_gama}).
As the surveys will cover up to full-sky area (SPHEREx full-sky, $\sim 40,000 \deg^2$) and our GAMA sample ($N \sim 30,000$ for $13 < i <18$) spans $\sim 200\deg^2$ in the sky, the surveys will cover up to $\lesssim 6$ millions of similarly bright galaxies across the sky and provide photo-$z$ values with sub-percent precision and accuracy.

The statistics from the deep dataset show comparable results even when we include fainter galaxies with $i\lesssim23$, reflecting the enhanced S/N of deep surveys (Table \ref{tab:stats_wide_deep}).
The values reveal that SPHEREx data show greater improvements compared to 7DS data when comparing the results between the wide and deep datasets.
SPHEREx data exhibits about 10 to 27 times smaller scatter, while 7DS data shows about 4 to 10 times improvement in scatter.
When we compare the statistics from the deep SPHEREx and 7DS datasets, SPHEREx outperforms 7DS in photo-$z$ performance.
Although SPHEREx has shallower depth in its individual bandpasses, it benefits from approximately five times as many bandpasses as 7DS, as well as wavelength coverage that is about ten times broader.
These factors collectively enhance the photo-$z$ measurements of SPHEREx.

The dense SED sampling also facilitates robust characterization of continuum features, making the photo-$z$ estimation more reliable for passive galaxies compared to star-forming galaxies. 
When we analyze the statistics as a function of rest-frame $B-V$ colors, the statistics are slightly better for redder galaxies consistently across most survey configurations.

Our result is generally consistent with \cite{feder24} using the SPHEREx, DECaLS \citep{dey19} \textit{g}\textit{r}\textit{z}, and WISE W1/W2 synthetic photometry.
They reported $\sigma = 0.0091$, $b = -0.0012$, and $\eta = 0.026$ for galaxies with $18.5 < W1 < 19.0$.
Our results demonstrate $\sigma = 0.0118\pm0.0011$, $b = -0.0002\pm0.0027$, and $\eta = 0.0082\pm0.0059$ for galaxies in the corresponding magnitude range ($19<i<20$) when using the SPHEREx in-house code, confirming consistency with previous results.
Note that the apple-to-apple comparison is impossible, as they calculated photo-$z$ performance statistics as a function of W1 magnitudes and incorporated broadband fluxes.

\cite{ko25} also conducted photo-$z$ measurement tests using mock 7DS catalog based on EL-COSMOS SEDs \citep{saito20}.
They estimated $\sigma = 0.004$, $b = -0.033$, and $\eta = 0.016$ for WTS within the galaxies with $19\leq m_{\rm 625} < 20$.
Our photo-$z$ measurements return $\sigma = 0.0053\pm0.0005$, $b = 0.0010\pm0.0008$, and $\eta = 0.0000\pm0.0000$ for WTS within the galaxies with $18.5 < i < 19.5$, which roughly corresponds to $19\leq m_{\rm 625} < 20$.
Therefore, our results are also consistent with \cite{ko25}, although the direct comparison is hard to achieve due to the difference in the sample and the survey parameters.

This level of photo-$z$ measurement performance is hardly achieved in conventional broadband wide-field surveys.
For example, \cite{tanaka18} tested the photo-$z$ estimations using Hyper Supreme-Cam Subaru Strategic Program (HSC-SSP) data with multiple ML-based and template-fitting methods.
Using five broadband photometric data ($g$, $r$, $i$, $z$, $y$), they achieved $\sigma \sim 0.02$, $b \sim 0.002$, and $\eta \sim 0.05$ for galaxies with $i\sim20$ (see Table 2 of \cite{tanaka18}). 
As they used deeper data than our wide dataset with a 5$\sigma$ point source depth of $r\sim 26$, the measured statistics remained similar down to $i \sim 23$.
However, the measured statistics from \cite{tanaka18} do not improve for brighter galaxies with $i \lesssim 19$, for which our results still show some improvements compared to galaxies with $i\sim20$.
The HSC-SSP provides deeper data compared to other surveys like SDSS \citep{york00_sdss, dominguez22_sdssdr17} or Pan-STARRS1 \citep{chambers16}.
Therefore, we can expect more improvements between our results and the photo-$z$ values based on these data, although it is not straightforward to directly compare the results.

\cite{ko25} also conducted a comparison between the photo-$z$ performance of the Pan-STARRS1 survey and mock 7DS data using EAZY \citep{brammer08}.
They also found that the measured $\eta$ and $\sigma$ are significantly better when they use 7DS data.
They found that photo-$z$ values from Pan-STARRS1 can deviate from their true redshifts even when the target is much brighter ($m_r < 20$) than the limiting magnitudes of the survey ($\lesssim 22.5$, c.f. Figure \ref{fig:depth}).

Furthermore, when \cite{ko25} combined the 7DS WTS and Pan-STARRS1 data for photo-$z$ measurement, they found improvements by a factor of two in the statistics for the faint targets ($m_{625} > 22$).
They interpreted it as the complement of the signal by relatively deeper PS1 surveys, although the catastrophic failures remain unchanged.
This result also demonstrates the potential synergy between the medium-band surveys and deeper broadband surveys like LSST \citep{ivezic19}, DESI \citep{dey19}, and Euclid \citep{laureijs11}.

\subsection{Synergy between SPHEREx and 7DS} \label{sec:dis_synergy}
The photo-$z$ statistics are persistently better in the combined dataset than in the SPHEREx-only or 7DS-only dataset, demonstrating the synergy between SPHEREx and 7DS in photo-$z$ estimations.
Figure \ref{fig:stats_gama} and \ref{fig:stats_cos} clearly show that the statistics of photo-$z$ measurements are the best for the combined data (the last row of the figures) compared to others, for most of the photo-$z$ methods.
The difference is more distinct for the galaxies within the fainter magnitude bins.
For example, when we compare the results from SPHEREx-only and SPHEREx + 7DS, Figure \ref{fig:stats_cos} shows that the combined dataset shows $\sim5$ times lower $\sigma$, $\sim80$ times lower $|b|$, and $\sim8$ times lower $\eta$ for the galaxies with $i\sim21.5$.
On the other hand, when we compare the results from 7DS-only and combined SPHEREx + 7DS data, the combined dataset shows comparable $\sigma$, but shows $\sim 7$ times lower $|b|$ and $\sim4$ times lower $\eta$ for the same magnitude bin.

The results imply that SPHEREx data contribute to photo-$z$ estimates, especially for $b$, even at the magnitude bins below the 5$\sigma$ depths of the SPHEREx full-sky survey within the wide dataset.
This is because the combination of SPHEREx and 7DS enhances the wavelength coverage.
The combination thus can lift the degeneracy between the color and redshift and reduce confusion from the low S/N data, making the photo-$z$ measurements better.
This is consistent with previous results \citep{feder24, ko25}, which showed that combining ancillary data with the survey spanning the wavelength range improves the photo-$z$ performances.

Figure \ref{fig:zz_1823} and \ref{fig:stats_cos} show that the improvement of the results from the SPHEREx to the SPHEREx + 7DS results is more pronounced compared to the difference between the 7DS and SPHEREx + 7DS data.
This difference is due to the higher S/N of WTS data, especially around galaxies with $i\sim21$, which are between the depth of WTS and the SPHEREx full-sky survey.
For the brighter galaxies in which WTS and the SPHEREx full-sky survey have enough signal, the statistics from SPHEREx and 7DS data become more similar (see Figure \ref{fig:zz_1318}).

\cite{feder24} noted that the forecasted photo-$z$ performance of SPHEREx already meets its planned science requirements.
Therefore, our results demonstrate the potential of the combined 7DS and SPHEREx data that can yield contributions to the planned sciences beyond the expected levels.
For example, the improved photo-$z$ estimations can help probe the large-scale structures and clustering of the galaxies.
We can also use this distance information for many science cases related to multi-messenger astronomy, like statistically inferring host galaxies of gravitational wave events within their localization area by matching objects and redshift peaks in galaxy catalogs \citep{singer16, dalya18} or that have high probabilities of hosting the event \citep{del_pozzo12, gair23, jeong24}.

The combination of SPHEREx and 7DS also reduces the dependence of photo-$z$ performance on the true redshifts.
Figure \ref{fig:z_stat} shows general improvements of the statistics in the combined dataset when we calculate the statistics of the galaxies within the true redshifts.
The statistics from SPHEREx data show lower performance compared to SPHEREx + 7DS data, especially for the galaxies with low redshifts.
The 7DS statistics show better results at low redshift, but it also has a larger $b$ compared to the combined data.
The longer wavelength coverage of the combined dataset can detect more spectral features and prevent confusion of the spectral features.
For example, low-redshift galaxies will lack some spectral features useful for identifying photo-$z$ values like the 4000\,\r{A} break or strong H$\alpha$ line in the SPHEREx data.
On the other hand, SPHEREx can capture 1.6$\,\mu$m bump and some PAH emissions, complementing the photo-$z$ estimation.

\subsection{Comparison between the Methods} \label{sec:dis_comparison}
The performance of the photo-$z$ measurement is also dependent on the photo-$z$ methods.
In Figure \ref{fig:zz_1318} and \ref{fig:zz_1823}, we plot the results from different methods in the different columns.
From Figure \ref{fig:stats_gama} to \ref{fig:z_stat}, we distinguish the methods with different colors.
Note that some differences are marginal, as only a few deviations of photo-$z$ measurements can yield this level of variation in statistics.
Therefore, we focus on our discussion on the differences between methods that persistently appear across multiple sets of input data.
In the discussion, we aim to offer insights for refining photo-$z$ measurement strategies using SPHEREx, 7DS, or a combined dataset.

SPHEREx in-house code (\citealp{stickley16, feder24}, Section \ref{sec:method_inhouse}) shows better results compared to other methods for the SPHEREx-only and SPHEREx + 7DS datasets in many cases.
This result is also evident in Figures \ref{fig:stats_gama} and \ref{fig:stats_cos}, where the orange lines representing the in-house code consistently exhibit superior statistical performance in the second row (SPHEREx wide data), especially for the fainter galaxies with $i \gtrsim 21$.
However, the code shows worse $b$ and $\eta$ for 7DS-only data and worse $\eta$ for SPHEREx + 7DS data compared to other methods for faint galaxies with $i\gtrsim20$.

When we compare the SPHEREx in-house code with LePhare, Figure \ref{fig:stats_gama} to \ref{fig:z_stat} shows that they have distinct trends of statistics as a function of magnitudes and redshifts from LePhare. 
For example, Figure \ref{fig:zz_1823} shows that LePhare and SPHEREx in-house code have notable differences for galaxies with $22 < i < 23$ when using 7DS-only data.
SPHEREx in-house code shows larger $b$, $\sigma$, and $\eta$ in this magnitude bin (see the 3rd row of Figure \ref{fig:stats_cos}).

LePhare and SPHEREx in-house code are different in the application of dust extinction laws, although the codes share a similar template fitting algorithm and the same template spectra.
As we noted in Section \ref{sec:method_inhouse}, the in-house code incorporates the dust extinction laws into the model grid.
We do not apply the additional dust extinction to LePhare (see Section \ref{sec:method_lephare}).
The dust extinction laws in the SPHEREx model introduce additional flexibility to the template spectra for the fitting.
Although LePhare applies a coarser grid size than the in-house code, we confirmed that the different grid sizes (0.002 v.s. 0.005) do not significantly affect the photo-$z$ performance.

The lower performance of the SPHEREx in-house code for 7DS data compared to LePhare can be attributed to the relative difficulty of constraining the dust extinction in the 7DS wavelength range alone.
The additional flexibility from the dust extinctions introduced degeneracy with the redshifts.
On the other hand, broad NIR wavelength coverage contains more continuum features like 1.6$\,\mu m$ bump and in long wavelength the spectral features are less susceptible to the extinction, making the redshift and extinction less degenerate to each other.
This result shows the importance of the treatment of dust extinctions on photo-$z$ estimations and wide wavelength coverage to constrain the dust extinctions. 
Both methods show similar performances for the brighter galaxies, implying that this effect is more important for fainter galaxies.

The results from EAZY and eazy-py are more similar to each other than LePhare and the SPHEREx in-house code.
They have some differences in the redshift grid, the detailed prescription of magnitude priors, the amplitude of constant template error (5\,\% for EAZY, 1\,\% for eazy-py), and the applied sets of templates. 
However, they share many aspects like treatment of dust extinction or application of magnitude prior and the template fitting algorithm, resulting in very similar statistics.
Considering that the operation time for the original EAZY code is much faster than eazy-py, when needing photo-$z$ estimations for large data, the EAZY code will be more appropriate to apply.
On the other hand, when we need more flexibility, eazy-py code will be more recommendable.

Since we construct a magnitude-redshift prior for EAZY using COSMOS galaxies while employing the built-in priors for eazy-py, the statistics from EAZY may be overestimated for COSMOS galaxies.
However, because other configurations also differ between EAZY and eazy-py, it is not straightforward to conclude that the prior alone causes the observed differences in results.
Nevertheless, we emphasize that SPHEREx and 7DS surveys require more optimized and independent priors to achieve better photo-$z$ estimations.

\subsubsection{Probability Density Functions} \label{sec:dis_pdf}

\begin{figure}[h!]
\includegraphics[width=0.47\textwidth]{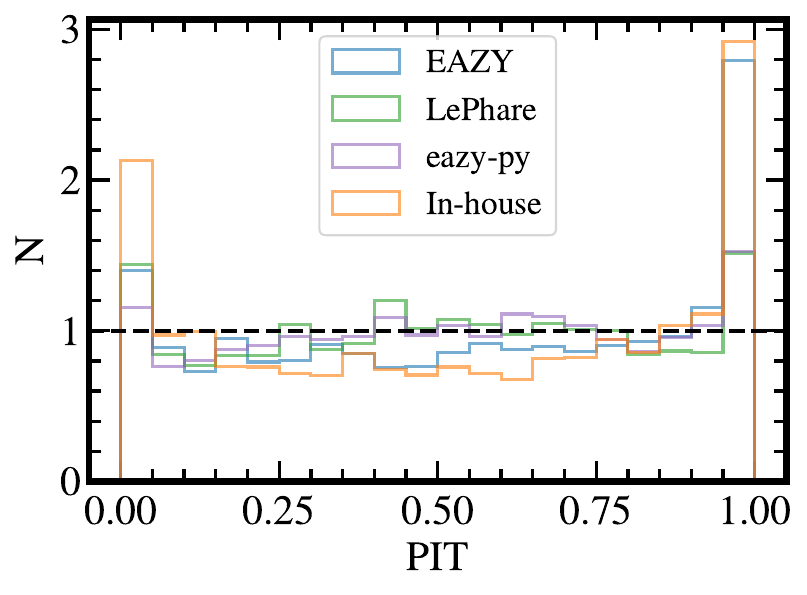}
\caption{The histogram of calculated PIT (Eq. \ref{eq:pit}) from the PDFs of different template fitting methods. We used COSMOS galaxies in the SPHEREx + 7DS wide dataset with $18<i<21$.}
\label{fig:pit}
\end{figure}

The template fitting codes provide probability density functions (PDFs). 
They are a valuable measure for checking the sanity and credibility of the fitting.
The codes also measure the confidence interval of the photo-z values using the PDFs \citep{arnouts99, brammer08, stickley16, brammer21}.
In this section, we compare the PDFs of the four template fitting codes by the metric named PIT (Probability Integral Transform, \citealp{polsterer16}). We calculated the PIT of the samples using the following equation from \cite{tanaka18}.
\begin{equation}
    \text{PIT}(z_{true}) = \int^{z_{true}}_0 P(z)dz.
    \label{eq:pit}
\end{equation}

We calculate the PIT values for COSMOS galaxies with $18<i<21$ applying no cuts on \texttt{FLAG\_ML}. 
The results based on the photo-$z$ estimates from the combined SPHEREx + 7DS data are shown in Figure \ref{fig:pit}.
The overall uniform distribution of the PIT values indicates that the PDFs are generally well calibrated, implying that the PDFs from the codes are reasonably well sampled and reflect the underlying uncertainties of the photo-$z$ estimates.
The peaks near PIT values of 0 and 1 correspond to outliers where the true redshifts deviate significantly from the redshift where the corresponding PDFs are located.
EAZY and the in-house code exhibit a mildly convex distribution, suggesting under-dispersed PDFs \citep{polsterer16}, whereas LePhare and eazy-py produce flatter PDFs, although the differences between the codes are not substantial.

\subsubsection{Comparison between Confidence Intervals} \label{sec:dis_confidence}

\begin{figure}[h!]
\includegraphics[width=0.47\textwidth]{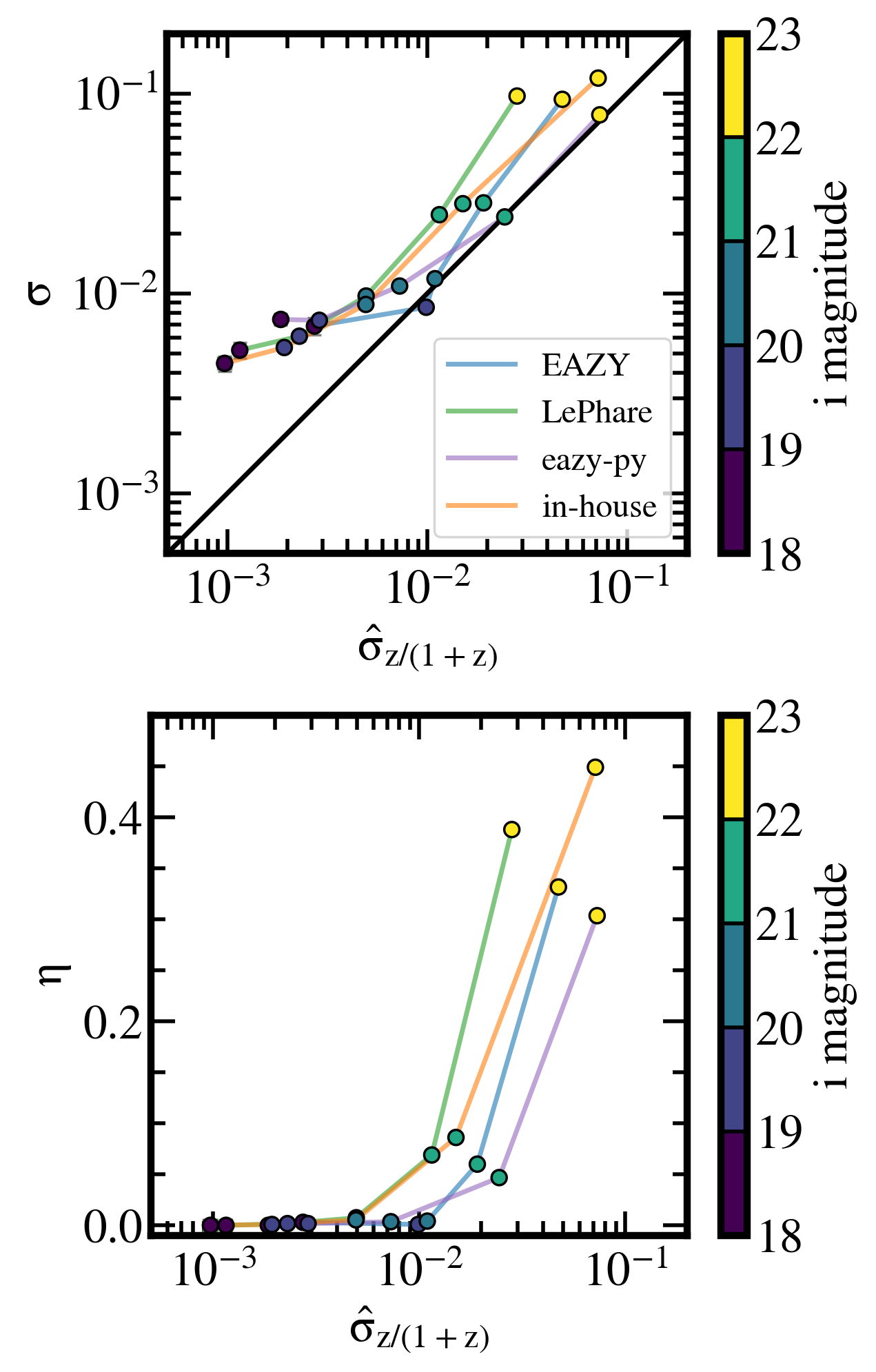}
\caption{The $\sigma$ as a function of the estimated redshift fractional uncertainties ($\hat{\sigma}_{z/(1+z)}$) from four different template fitting codes, binned by the magnitude of the sample. 
The color of the scatter shows the $i$-magnitude of each bin as noted in the colorbar.
We use the COSMOS galaxies with $18<i<23$ in the wide depth and SPHEREx + 7DS survey combination.
Most of the error bars in each point are invisible due to the markers.
}
\label{fig:sigma}
\end{figure}

We assess the credibility of the estimated confidence intervals of template fitting methods by comparing them with the real scatter between the photo-$z$s and the true redshift values.
The reliability of the error estimates is important, as in real observations, we do not know the true redshifts. 
For this analysis, we use the results obtained from the SPHEREx + 7DS dataset for COSMOS galaxies in the wide-depth sample, including the training sample as well.

In Figure \ref{fig:sigma}, we show the relation between the estimated redshift fractional uncertainties ($\hat{\sigma}_{z/(1+\hat{z})} = \hat{\sigma}_z/(1+\hat{z})^2$, where $\hat{\sigma}_z$ is the half of the reported 68\,\% confidence interval.) and real scatter in the photo-$z$ measurements, $\sigma$ ($\sigma_{\textrm{NMAD}}$ in this study), and the outlier fraction, $\eta$.
The intrinsic scatter in the photo-$z$ measurements, $\sigma$ exhibits an approximately linear relation with $\hat{\sigma}_{z/(1+\hat{z})}$.
However, the estimated errors generally underestimate the real scatter, $\sigma$.
This underestimation is more significant for brighter galaxies.

The underestimation of the confidence intervals in EAZY and the SPHEREx in-house code is consistent with the results from the PIT analysis.
In contrast, LePhare shows a flat to mildly convex distribution in the PIT distribution, implying overdispersed PDFs, yet has underestimated confidence intervals. 
We attribute this to the multi-peaked shape of the PDFs of some samples in LePhare.
The presence of multiple peaks broadens the overall PDF, while the confidence intervals derived from the $\chi^2$ statistics in LePhare are typically confined to one peak, leading to underestimation of the confidence intervals.

As noted in Section \ref{sec:method_eazypy}, eazy-py in this study applies a correction factor of 1.5 to the flux uncertainties, which results in more consistent estimated errors to the real photo-$z$ scatter compared to the other codes.
The scaling behaves as expected for the fainter sample, as shown in the upper panel of Figure \ref{fig:sigma}, because the sample for correcting flux errors is dominated by faint galaxies near the limiting magnitude of the data.
The PIT distribution from eazy-py (purple histogram in Figure \ref{fig:pit}) also shows more flatter distribution compared to other codes. 
This suggests that the treatment of correcting uncertainties should account for the magnitude or S/N.

The lower panel of Figure \ref{fig:sigma} shows the outlier fraction ($\eta$) as a function of $\hat{\sigma}_{z/(1+z)}$.
As expected, the outlier fraction becomes larger with the estimated errors for all the codes.
Generally, LePhare and in-house code show larger $\eta$ compared to EAZY and eazy-py.

\subsubsection{Notes on ML-Based Methods} \label{sec:dis_ml}
In Figure \ref{fig:zz_1823}, the results from HRF (Section \ref{sec:method_rf}, see also \citealp{kim25a}) show conceivably low scatter in the faintest magnitude bins for combined SPHEREx + 7DS and 7DS-only data.
However, the HRF shows larger scatter for bright samples, leading to relatively larger $\sigma$ for GAMA galaxies (c.f. Figure \ref{fig:zz_1318}) for SPHEREx-only and SPHEREx + 7DS datasets.

To understand the distinct trend of the statistics as a function of magnitudes, we investigate the photo-$z$ measurements using the HRF method in more detail.
As described in Section \ref{sec:method_rf}, HRF uses flux uncertainties as input features in addition to the flux values. 
We build another HRF model without flux uncertainties for the wide dataset to check the dependence of photo-$z$ measurement on the flux uncertainties.
The resulting photo-$z$ shows similar results to other methods for SPHEREx + 7DS and 7DS data, with increased absolute values of the statistics, especially for the faint magnitudes ($21 < i< 23$).
The results from SPHEREx data remain similar before and after removing flux uncertainties in the input features, but the HRF results are already similar to other methods for the dataset (see Figure \ref{fig:stats_cos}).

These results imply that the flux uncertainty directly influences the redshift measurements in the HRF method, especially for the objects with low S/N.
The flux uncertainty does not affect the redshift estimation directly in the SED fitting-based method, although a larger uncertainty leads to a smoother PDF and larger photo-$z$ uncertainty estimates.
Using flux uncertainty to infer photo-$z$ may be helpful as we utilize additional information, provided that the survey is homogeneous enough to establish a reliable relationship between flux uncertainties and redshift.
However, when the survey is inhomogeneous, the information from the flux uncertainty can be biased, resulting in biased photo-$z$ estimates depending on the training set.
Therefore, to utilize flux uncertainties in the model, we should consider additional information that can affect the flux uncertainties, like the number of visits and the quality of the data.

The HRF method generally underperforms for bright galaxies compared to other methods, as shown in Figures \ref{fig:zz_1318} and \ref{fig:stats_gama}.
It is likely to originate from the HRF model's limited capacity for extrapolating the inferred nonlinear relationships beyond the training set.
To make the HRF more accurate, we may need more training samples spanning wider redshifts and galaxy types.

The DNN method (Section \ref{sec:method_dnn}) shows a good performance for bright galaxies in these figures. 
Although the DNN method shows outstanding results for bright GAMA galaxies, the DNN method is difficult to apply to fainter targets with numerous non-detections (see Section \ref{sec:method_dnn}). 
Since the data from SPHEREx and 7DS surveys contain many bands that divide the signal more than broadband surveys, some non-detections are inevitable in the data.
Therefore, to utilize DNN throughout the survey, we need to develop more robust methods for handling non-detections.
Despite that, DNN will add reliable photo-$z$ estimates to many bright galaxies.

\section{Conclusion} \label{sec:conclusion}
In this paper, we predict the photo-$z$ performances of SPHEREx, 7DS, and combined SPHEREx + 7DS data by measuring photo-$z$ values from SPHEREx and 7DS mock catalogs using six different photo-$z$ methods. 
We summarize the main findings of this study as follows:

\begin{itemize}
  \item We construct the mock catalogs of SPHEREx and 7DS surveys by convolving mock SEDs from \cite{feder24} with SPHEREx and 7DS filter responses and considering observing conditions. 

  \item We test photo-$z$ measurement using EAZY, eazy-py, LePhare, SPHEREx in-house code, RF, and DNN. We provide results mainly for SPHEREx full-sky + 7DS WTS (the wide dataset) and SPHEREx deep + 7DS IMS (the deep dataset).
  
  \item   For bright galaxies with $13< i< 18$ in the wide dataset (SPHEREx full-sky + 7DS WTS), the photo-$z$ values show $\sigma \lesssim 0.005$, $b \lesssim 0.001$, and $\eta \lesssim 0.002$ when we use the combined SPHEREx and 7DS data.
  We measure $\sigma \lesssim 0.01$, $b \lesssim 0.003$, and $\eta \lesssim 0.01$ for galaxies with $18<i<21$ with the same survey combination. 
  These results show significant improvements compared to the literature broadband-based photo-$z$ results, especially for bright targets with enough S/N for the SPHEREx and 7DS bands.

  \item The deep dataset (SPHEREx deep + 7DS IMS) shows comparable performance for the galaxies that are up to two magnitudes fainter than those in the wide dataset.

  \item The SPHEREx + 7DS combined dataset shows better results for almost all galaxies and methods, showcasing the synergy between the surveys. 
  The photo-$z$ values from the combined dataset also show lower dependence on the magnitudes and true redshifts of the galaxies.

  \item The methods show differences in photo-$z$ performances, especially for the faint galaxies. 
  The SPHEREx in-house code generally shows the best results for SPHEREx-only data and SPHEREx + 7DS data for galaxies with $i\lesssim 22$.
  
  \item Although sharing the template fitting algorithm, LePhare and SPHEREx in-house pipeline show different trends of statistics as a function of magnitudes and redshifts.
  The difference mainly comes from the application of dust extinction laws in the SPHEREx in-house code.

  \item We show from the PIT analysis that the PDFs of the four template-fitting codes are generally well calibrated, but with a mild sign of underdispersed PDF for in-house and EAZY.
  We also examined the estimated confidence intervals, which scale with the real scatter of the photo-$z$ estimates, and found that boosting the flux uncertainties helps correct the tendency to underestimate the real scatter in the photo-$z$ values.
  
  \item RF shows noticeably better results for faint galaxies for the combined SPHEREx + 7DS and 7DS-only data for faint galaxies with $i >22$. 
  However, this method exploits flux uncertainties for photo-$z$ estimation, which can lead to bias for real data that can have some inhomogeneities.
  
  \item DNN shows better statistics for bright galaxies, but its applicability is limited for fainter targets with many non-detections.
  
\end{itemize}

Our results demonstrate the capability of SPHEREx and 7DS to provide a large number of precise photometric redshifts.
By comparing several widely used photo-$z$ methods, we examined the main factors that influence photo-$z$ performance and highlighted the complementary nature of the two surveys.

The aim of this work is not to optimize individual codes or to identify a preferred method, but rather to assess how representative photo-$z$ techniques respond to simulated SPHEREx and 7DS data in preparation for the upcoming observations. Given the unique characteristics of SPHEREx, dense narrow-band sampling, and the anticipated synergy with 7DS, the combined dataset shows improved performance compared to either survey alone. This study, therefore, provides a baseline expectation for photo-$z$ measurements from future SPHEREx and 7DS observations.

\begin{acknowledgement}
We are grateful to the anonymous referee for giving valuable comments to improve the content of this paper.
This work was supported by the National Research Foundation of Korea (NRF) grant, No. 2021M3F7A1084525, funded by the Korean government (MSIT).
HSH acknowledges the support of Samsung Electronics Co., Ltd. (Project Number IO220811-01945-01), the NRF grant funded by the Korean government (MSIT), NRF-2021R1A2C1094577, and Hyunsong Educational \& Cultural Foundation. 
Y. K. was supported by the NRF grant funded by the Korean government (MSIT) (No. 2021R1C1C2091550). 
B. L. is supported by the NRF grant funded by the Korea government(MSIT) (No. NRF-2022R1C1C1008695).
M. K. was supported by the NRF grant funded by the Korean government (MSIT) (No. RS-2024-00347548).
SL acknowledges support from the NRF grant (RS-2025-00573214) funded by the Korean government(MSIT).
S. K. was supported by the NRF grant funded by the Korean government (MSIT) (No. 2022R1C1C2005539).
The work of HB was supported by the Basic Science Research Program through the NRF funded by the Ministry of Education (RS-2025-25403440).
D.K. acknowledges the support by the NRF grant (No. 2021R1C1C1013580) funded by the Korean government (MSIT).
J.H.K. acknowledges the Institute of Information \& Communications Technology Planning \& Evaluation (IITP) grant, No. RS-2021-II212068 funded by the Korean government (MSIT).
\end{acknowledgement}

\bibliographystyle{aa}
\bibliography{main.bib}

\end{document}